\input harvmac
\noblackbox


\def\thp{{\theta'}}
\def\thpl{{\theta'_L}}
\def\thpr{{\theta'_R}}
\def\thpg{{\theta'_G}}
\def\thplb{{\theta'_{LB}}}
\def\thprb{{\theta'_{RB}}}

\def\IZ{\relax\ifmmode\mathchoice
{\hbox{\cmss Z\kern-.4em Z}}{\hbox{\cmss Z\kern-.4em Z}}
{\lower.9pt\hbox{\cmsss Z\kern-.4em Z}} {\lower1.2pt\hbox{\cmsss
Z\kern-.4em Z}}\else{\cmss Z\kern-.4em Z}\fi}
\def\IB{\relax{\rm I\kern-.18em B}}
\def\IC{{\relax\hbox{\kern.3em{\cmss I}$\kern-.4em{\rm C}$}}}
\def\ID{\relax{\rm I\kern-.18em D}}
\def\IE{\relax{\rm I\kern-.18em E}}
\def\IF{\relax{\rm I\kern-.18em F}}
\def\IG{\relax\hbox{$\inbar\kern-.3em{\rm G}$}}
\def\IGa{\relax\hbox{${\rm I}\kern-.18em\Gamma$}}
\def\IH{\relax{\rm I\kern-.18em H}}
\def\II{\relax{\rm I\kern-.18em I}}
\def\IK{\relax{\rm I\kern-.18em K}}
\def\IP{\relax{\rm I\kern-.18em P}}

\font\cmss=cmss10 \font\cmsss=cmss10 at 7pt
\def\IR{\relax{\rm I\kern-.18em R}}

\def\frac#1#2{{#1 \over #2}}

\def\OL#1{ \kern1pt\overline{\kern-1pt#1
   \kern-1pt}\kern1pt }

\lref\BirrellIX{ N.~D.~Birrell and P.~C.~Davies, ``Quantum Fields
In Curved Space,''}

\lref\KachruAW{ S.~Kachru, R.~Kallosh, A.~Linde and S.~P.~Trivedi,
``De Sitter vacua in string theory,'' arXiv:hep-th/0301240.
}

\lref\KachruHE{ S.~Kachru, M.~B.~Schulz and S.~Trivedi, ``Moduli
stabilization from fluxes in a simple IIB orientifold,''
arXiv:hep-th/0201028.
}

\lref\SusskindIF{
L.~Susskind, L.~Thorlacius and J.~Uglum,
``The Stretched horizon and black hole complementarity,''
Phys.\ Rev.\ D {\bf 48}, 3743 (1993)
[arXiv:hep-th/9306069].
}

\lref\KachruNS{ S.~Kachru, X.~Liu, M.~B.~Schulz and S.~P.~Trivedi,
``Supersymmetry changing bubbles in string theory,''
arXiv:hep-th/0205108.
}

\lref\MaloneyRR{ A.~Maloney, E.~Silverstein and A.~Strominger,
``De Sitter space in noncritical string theory,''
arXiv:hep-th/0205316.
}

\lref\SilversteinXN{ E.~Silverstein, ``(A)dS backgrounds from
asymmetric orientifolds,'' arXiv:hep-th/0106209.
}

\lref\ItzhakiDD{ N.~Itzhaki, J.~M.~Maldacena, J.~Sonnenschein and
S.~Yankielowicz, ``Supergravity and the large N limit of theories
with sixteen  supercharges,'' Phys.\ Rev.\ D {\bf 58}, 046004
(1998) [arXiv:hep-th/9802042].
}

\lref\BoussoXA{ R.~Bousso and J.~Polchinski, ``Quantization of
four-form fluxes and dynamical neutralization of the  cosmological
constant,'' JHEP {\bf 0006}, 006 (2000) [arXiv:hep-th/0004134].
}

\lref\AcharyaKV{ B.~S.~Acharya, ``A moduli fixing mechanism in M
theory,'' arXiv:hep-th/0212294.
}

\lref\SusskindKW{ L.~Susskind, ``The anthropic landscape of string
theory,'' arXiv:hep-th/0302219.
}

\lref\DouglasUM{ M.~R.~Douglas, ``The statistics of string / M
theory vacua,'' arXiv:hep-th/0303194.
}

\lref\GiddingsYU{ S.~B.~Giddings, S.~Kachru and J.~Polchinski,
``Hierarchies from fluxes in string compactifications,'' Phys.\
Rev.\ D {\bf 66}, 106006 (2002) [arXiv:hep-th/0105097].
}

\lref\KachruGS{ S.~Kachru, J.~Pearson and H.~Verlinde,
``Brane/flux annihilation and the string dual of a
non-supersymmetric  field theory,'' JHEP {\bf 0206}, 021 (2002)
[arXiv:hep-th/0112197].
}

\lref\AcharyaDZ{ B.~S.~Acharya and C.~Vafa, ``On domain walls of N
= 1 supersymmetric Yang-Mills in four dimensions,''
arXiv:hep-th/0103011.
}

\lref\Andreas{A. Karch, work in progress}

\lref\GukovYA{ S.~Gukov, C.~Vafa and E.~Witten, ``CFT's from
Calabi-Yau four-folds,'' Nucl.\ Phys.\ B {\bf 584}, 69 (2000)
[Erratum-ibid.\ B {\bf 608}, 477 (2001)] [arXiv:hep-th/9906070].
}

\lref\KrausHV{ P.~Kraus, F.~Larsen and S.~P.~Trivedi, ``The
Coulomb branch of gauge theory from rotating branes,'' JHEP {\bf
9903}, 003 (1999) [arXiv:hep-th/9811120].
}

\lref\FengIF{ J.~L.~Feng, J.~March-Russell, S.~Sethi and
F.~Wilczek, ``Saltatory relaxation of the cosmological constant,''
Nucl.\ Phys.\ B {\bf 602}, 307 (2001) [arXiv:hep-th/0005276].
}

\lref\FreyHF{ A.~R.~Frey and J.~Polchinski, ``N = 3 warped
compactifications,'' Phys.\ Rev.\ D {\bf 65}, 126009 (2002)
[arXiv:hep-th/0201029].
}

\lref\GibbonsMU{ G.~W.~Gibbons and S.~W.~Hawking, ``Cosmological
Event Horizons, Thermodynamics, And Particle Creation,'' Phys.\
Rev.\ D {\bf 15}, 2738 (1977).
}

\lref\ColemanAW{ S.~R.~Coleman and F.~De Luccia, ``Gravitational
Effects On And Of Vacuum Decay,'' Phys.\ Rev.\ D {\bf 21}, 3305
(1980).
}

\lref\MaldacenaRE{ J.~M.~Maldacena, ``The large N limit of
superconformal field theories and supergravity,'' Adv.\ Theor.\
Math.\ Phys.\  {\bf 2}, 231 (1998) [Int.\ J.\ Theor.\ Phys.\  {\bf
38}, 1113 (1999)] [arXiv:hep-th/9711200].
}

\lref\BrownKG{ J.~D.~Brown and C.~Teitelboim, ``Neutralization Of
The Cosmological Constant By Membrane Creation,'' Nucl.\ Phys.\ B
{\bf 297}, 787 (1988).
}

\lref\BanksNM{ T.~Banks, ``Heretics of the false vacuum:
Gravitational effects on and of vacuum  decay. II,''
arXiv:hep-th/0211160.
}

\lref\freyetal{A. Frey, M. Lippert, and B. Williams, to appear}

\lref\KrausHV{ P.~Kraus, F.~Larsen and S.~P.~Trivedi, ``The
Coulomb branch of gauge theory from rotating branes,'' JHEP {\bf
9903}, 003 (1999) [arXiv:hep-th/9811120].
}

\lref\HorowitzNW{ G.~T.~Horowitz and J.~Polchinski, ``A
correspondence principle for black holes and strings,'' Phys.\
Rev.\ D {\bf 55}, 6189 (1997) [arXiv:hep-th/9612146].
}

\lref\FlanaganJP{ E.~E.~Flanagan, D.~Marolf and R.~M.~Wald,
``Proof of Classical Versions of the Bousso Entropy Bound and of
the Generalized Second Law,'' Phys.\ Rev.\ D {\bf 62}, 084035
(2000) [arXiv:hep-th/9908070].
}

\lref\SusskindWS{ L.~Susskind, ``Some Speculations About Black
Hole Entropy In String Theory,'' arXiv:hep-th/9309145.
}

\lref\BoussoXY{ R.~Bousso, ``A Covariant Entropy Conjecture,''
JHEP {\bf 9907}, 004 (1999) [arXiv:hep-th/9905177].
}

\lref\SusskindDQ{ L.~Susskind and E.~Witten, ``The holographic
bound in anti-de Sitter space,'' arXiv:hep-th/9805114.
}

\lref\StromingerSH{ A.~Strominger and C.~Vafa, ``Microscopic
Origin of the Bekenstein-Hawking Entropy,'' Phys.\ Lett.\ B {\bf
379}, 99 (1996) [arXiv:hep-th/9601029].
}

\lref\DysonNT{ L.~Dyson, J.~Lindesay and L.~Susskind, ``Is there
really a de Sitter/CFT duality,'' JHEP {\bf 0208}, 045 (2002)
[arXiv:hep-th/0202163].
}

\lref\GutperleAI{ M.~Gutperle and A.~Strominger, ``Spacelike
branes,'' JHEP {\bf 0204}, 018 (2002) [arXiv:hep-th/0202210].
}

\lref\SenIN{ A.~Sen, ``Tachyon matter,'' JHEP {\bf 0207}, 065
(2002) [arXiv:hep-th/0203265].
}

\lref\SusskindSM{ L.~Susskind and J.~Uglum, ``Black hole entropy
in canonical quantum gravity and superstring theory,'' Phys.\
Rev.\ D {\bf 50}, 2700 (1994) [arXiv:hep-th/9401070].
}

\lref\RandallTG{ L.~Randall, V.~Sanz and M.~D.~Schwartz,
``Entropy-area relations in field theory,'' JHEP {\bf 0206}, 008
(2002) [arXiv:hep-th/0204038].
}

\lref\HawkingDA{ S.~Hawking, J.~M.~Maldacena and A.~Strominger,
``DeSitter entropy, quantum entanglement and AdS/CFT,'' JHEP {\bf
0105}, 001 (2001) [arXiv:hep-th/0002145].
}

\lref\MaldacenaIH{ J.~M.~Maldacena and A.~Strominger,
``Statistical entropy of de Sitter space,'' JHEP {\bf 9802}, 014
(1998) [arXiv:gr-qc/9801096].
}

\lref\KachruYS{ S.~Kachru and E.~Silverstein, ``4d conformal
theories and strings on orbifolds,'' Phys.\ Rev.\ Lett.\  {\bf
80}, 4855 (1998) [arXiv:hep-th/9802183].
}

\lref\BanksFE{ T.~Banks, ``Cosmological breaking of supersymmetry
or little Lambda goes back to  the future. II,''
arXiv:hep-th/0007146.
}

\lref\StromingerBR{ A.~Strominger and D.~Thompson, ``A quantum
Bousso bound,'' arXiv:hep-th/0303067.
}

\lref\BoussoJU{ R.~Bousso, ``The holographic principle,'' Rev.\
Mod.\ Phys.\  {\bf 74}, 825 (2002) [arXiv:hep-th/0203101].
}

\input epsf
\noblackbox
\newcount\figno
\figno=0
\def\fig#1#2#3{
\par\begingroup\parindent=0pt\leftskip=1cm\rightskip=1cm\parindent=0pt
\baselineskip=11pt \global\advance\figno by 1 \midinsert
\epsfxsize=#3 \centerline{\epsfbox{#2}} \vskip 12pt {\bf Fig.\
\the\figno: } #1\par
\endinsert\endgroup\par
}
\def\figlabel#1{\xdef#1{\the\figno}}

\lref\fischler{ W. Fischler, unpublished. }

\lref\GukovYA{ S.~Gukov, C.~Vafa and E.~Witten, ``CFT's from
Calabi-Yau four-folds,'' Nucl.\ Phys.\ B {\bf 584}, 69 (2000)
[Erratum-ibid.\ B {\bf 608}, 477 (2001)] [arXiv:hep-th/9906070].
}

\lref\GiddingsYU{ S.~B.~Giddings, S.~Kachru and J.~Polchinski,
``Hierarchies from fluxes in string compactifications,''
arXiv:hep-th/0105097.
}

\lref\BeckerGJ{ K.~Becker and M.~Becker, ``M-Theory on
Eight-Manifolds,'' Nucl.\ Phys.\ B {\bf 477}, 155 (1996)
[arXiv:hep-th/9605053].
}

\lref\PolchinskiSM{ J.~Polchinski and A.~Strominger, ``New Vacua
for Type II String Theory,'' Phys.\ Lett.\ B {\bf 388}, 736 (1996)
[arXiv:hep-th/9510227].
}

\lref\GukovYA{ S.~Gukov, C.~Vafa and E.~Witten,
Nucl.\ Phys.\ B {\bf 584}, 69 (2000) [Erratum-ibid.\ B {\bf 608},
477 (2001)] [arXiv:hep-th/9906070].
}

\lref\SilversteinXN{E.~Silverstein,``(A)dS backgrounds from
asymmetric orientifolds,''arXiv:hep-th/0106209. }
\lref\BrownDD{J.~D.~Brown and C.~Teitelboim,``Dynamical
Neutralization Of The Cosmological Constant,''Phys.\ Lett.\ B {\bf
195}, 177 (1987). } \lref\AbbottQF{L.~F.~Abbott,``A Mechanism For
Reducing The Value Of The Cosmological Constant,''Phys.\ Lett.\ B
{\bf 150}, 427 (1985).} \lref\BrownKG{J.~D.~Brown and
C.~Teitelboim,``Neutralization Of The Cosmological Constant By
Membrane Creation,''Nucl.\ Phys.\ B {\bf 297}, 787 (1988). }
\lref\HawkingMY{S.~W.~Hawking and I.~G.~Moss,``Fluctuations In The
Inflationary Universe,''Nucl.\ Phys.\ B {\bf 224}, 180 (1983).}
\lref\ColemanAW{S.~R.~Coleman and F.~De Luccia,``Gravitational
Effects On And Of Vacuum Decay,''Phys.\ Rev.\ D {\bf 21}, 3305
(1980).} \lref\BoussoXA{R.~Bousso and J.~Polchinski,``Quantization
of four-form fluxes and dynamical neutralization of the
cosmological constant,''JHEP {\bf 0006}, 006
(2000)[arXiv:hep-th/0004134].}
\lref\deAlwisPR{ S.~P.~de Alwis, J.~Polchinski and R.~Schimmrigk,
``Heterotic Strings With Tree Level Cosmological Constant,''
Phys.\ Lett.\ B {\bf 218}, 449 (1989).
} \lref\psstw{ J.~Preskill, P.~Schwarz, A.~D.~Shapere, S.~Trivedi
and F.~Wilczek, ``Limitations on the statistical description of
black holes,'' Mod.\ Phys.\ Lett.\ A {\bf 6}, 2353 (1991).}
\lref\jmls{J.~M.~Maldacena and L.~Susskind, ``D-branes and Fat
Black Holes,'' Nucl.\ Phys.\ B {\bf 475}, 679 (1996)
[arXiv:hep-th/9604042].}
\lref\PolyakovJU{ A.~M.~Polyakov, ``The wall of the cave,'' Int.\
J.\ Mod.\ Phys.\ A {\bf 14}, 645 (1999) [arXiv:hep-th/9809057].
}
\lref\FengIF{ J.~L.~Feng, J.~March-Russell, S.~Sethi and
F.~Wilczek, ``Saltatory relaxation of the cosmological constant,''
Nucl.\ Phys.\ B {\bf 602}, 307 (2001) [arXiv:hep-th/0005276].
}
\lref\ChamseddineQU{ A.~H.~Chamseddine, ``A Study of noncritical
strings in arbitrary dimensions,'' Nucl.\ Phys.\ B {\bf 368}, 98
(1992).
}

\lref\KachruHE{ S.~Kachru, M.~Schulz and S.~Trivedi, ``Moduli
stabilization from fluxes in a simple IIB orientifold,''
arXiv:hep-th/0201028.
}
\lref\FreyHF{ A.~R.~Frey and J.~Polchinski, ``N = 3 warped
compactifications,'' arXiv:hep-th/0201029.
}
\lref\DysonNT{L.~Dyson, J.~Lindesay and L.~Susskind, ``Is there
really a de Sitter/CFT duality,''arXiv:hep-th/0202163.
}
\lref\CrapsII{ B.~Craps, D.~Kutasov and G.~Rajesh, ``String
Propagation in the Presence of Cosmological Singularities,''
arXiv:hep-th/0205101.
} \lref\gp{ P.~Ginsparg and M.~J.~Perry, ``Semiclassical
Perdurance Of De Sitter Space,'' Nucl.\ Phys.\ B {\bf 222}, 245
(1983).}

\lref\gh{ G.~W.~Gibbons and C.~M.~Hull, ``de Sitter space from
warped supergravity solutions,'' arXiv:hep-th/0111072.}

\lref\StromingerPN{ A.~Strominger, ``The dS/CFT correspondence,''
JHEP {\bf 0110}, 034 (2001) [arXiv:hep-th/0106113].
}

\lref\chu{ C.~M.~Hull, ``de Sitter space in supergravity and M
theory,'' JHEP {\bf 0111}, 012 (2001) [arXiv:hep-th/0109213].}
\lref\fone{T.~Banks, ``Cosmological Breaking Of Supersymmetry?,''
Int.\ J.\ Mod.\ Phys.\ A {\bf 16}, 910 (2001).}
      \lref\ftwo{
R.~Bousso, ``Positive vacuum energy and the N-bound,'' JHEP {\bf
0011}, 038 (2000) [arXiv:hep-th/0010252].} \lref\per{ P.~Berglund,
T.~Hubsch and D.~Minic, ``de Sitter spacetimes from warped
compactifications of IIB string theory,'' arXiv:hep-th/0112079.}
\lref\vn{ K.~Pilch, P.~van Nieuwenhuizen and M.~F.~Sohnius, ``De
Sitter Superalgebras And Supergravity,'' Commun.\ Math.\ Phys.\
{\bf 98}, 105 (1985).} \lref\fre{ P.~Fre, M.~Trigiante and A.~Van
Proeyen, ``Stable de Sitter Vacua from N=2 Supergravity,''
arXiv:hep-th/0205119.} \lref\kallosh{ R.~Kallosh, ``N = 2
supersymmetry and de Sitter space,'' arXiv:hep-th/0109168.}
\lref\chamblin{ A.~Chamblin and N.~D.~Lambert, ``de Sitter space
from M-theory,'' Phys.\ Lett.\ B {\bf 508}, 369 (2001)
[arXiv:hep-th/0102159].} \lref\susskind{ L.~Susskind, ``Twenty
years of debate with Stephen,'' arXiv:hep-th/0204027.}
     \lref\jmas{
J.~M.~Maldacena and A.~Strominger, ``Statistical entropy of de
Sitter space,'' JHEP {\bf 9802}, 014 (1998)
[arXiv:gr-qc/9801096].}

\lref\StromingerJG{ A.~Strominger, ``The Inverse Dimensional
Expansion In Quantum Gravity,'' Phys.\ Rev.\ D {\bf 24}, 3082
(1981).
}

\lref\GatesCT{ S.~J.~Gates and B.~Zwiebach, ``Gauged N=4
Supergravity Theory With A New Scalar Potential,'' Phys.\ Lett.\ B
{\bf 123}, 200 (1983).
}
\lref\LindeSK{ A.~D.~Linde, ``Hard art of the universe creation
(stochastic approach to tunneling and baby universe formation),''
Nucl.\ Phys.\ B {\bf 372}, 421 (1992) [arXiv:hep-th/9110037].
}
\lref\RohmAQ{ R.~Rohm, ``Spontaneous Supersymmetry Breaking In
Supersymmetric String Theories,'' Nucl.\ Phys.\ B {\bf 237}, 553
(1984).
}

\lref\DasguptaSS{ K.~Dasgupta, G.~Rajesh and S.~Sethi, ``M theory,
orientifolds and G-flux,'' JHEP {\bf 9908}, 023 (1999)
[arXiv:hep-th/9908088].
}

\lref\KachruGS{ S.~Kachru, J.~Pearson and H.~Verlinde,
``Brane/flux annihilation and the string dual of a
non-supersymmetric  field theory,'' arXiv:hep-th/0112197.
}

\lref\raph{R. Bousso, ''Adventures in de Sitter space'',
hep-th/0205177.}

\lref\raphb{ R.~Bousso, O.~DeWolfe and R.~C.~Myers, ``Unbounded
entropy in spacetimes with positive cosmological constant,''
arXiv:hep-th/0205080.
.}

\lref\HalyoPX{
E.~Halyo,
``De Sitter entropy and strings,''
arXiv:hep-th/0107169.
}

\lref\BalasubramanianRB{ V.~Balasubramanian, P.~Ho\v{r}ava and
D.~Minic, ``Deconstructing de Sitter,'' JHEP {\bf 0105}, 043
(2001) [arXiv:hep-th/0103171].
}

\lref\GarrigaEF{ J.~Garriga and A.~Vilenkin, ``Recycling
universe,'' Phys.\ Rev.\ D {\bf 57}, 2230 (1998)
[arXiv:astro-ph/9707292].
}

\lref\simeon{S. Hellerman, private communication.}

\lref\WittenKN{
E.~Witten,
``Quantum gravity in de Sitter space,''
arXiv:hep-th/0106109.
}

\lref\GoheerVF{
N.~Goheer, M.~Kleban and L.~Susskind,
``The trouble with de Sitter space,''
arXiv:hep-th/0212209.
}

\lref\ParkQK{ M.~I.~Park, ``Statistical entropy of
three-dimensional Kerr-de Sitter space,'' Phys.\ Lett.\ B {\bf
440}, 275 (1998) [arXiv:hep-th/9806119].
}

\lref\ParkYW{ M.~I.~Park, ``Symmetry algebras in Chern-Simons
theories with boundary: Canonical  approach,'' Nucl.\ Phys.\ B
{\bf 544}, 377 (1999) [arXiv:hep-th/9811033].
}


\Title{\vbox{\baselineskip12pt\hbox{hep-th/0304220}
\hbox{SLAC-PUB-9717}\hbox{SU-ITP-03/08}}} {\vbox{
\centerline{D-Sitter Space:}
\bigskip
\centerline{Causal Structure, Thermodynamics, and Entropy}
 }}

\centerline{Michal Fabinger and Eva Silverstein}
\bigskip
\centerline{ \sl SLAC and Department of Physics}

 \centerline{ \sl Stanford University}

 \centerline{ \sl Stanford, CA 94305/94309}




\bigskip
\bigskip

\vskip .3in \centerline{\bf Abstract} {We study the entropy of
concrete de Sitter flux compactifications and deformations of them
containing D-brane domain walls.  We determine the relevant causal
and thermodynamic properties of these ``D-Sitter" deformations of
de Sitter spacetimes. We find a string scale correspondence point
at which the entropy localized on the D-branes (and measured by
probes sent from an observer in the middle of the bubble) scales
the same with large flux quantum numbers as the entropy of the
original de Sitter space, and at which Bousso's bound is saturated
by the D-brane degrees of freedom (up to order one coefficients)
for an infinite range of times. From the geometry of a static
patch of D-Sitter space and from basic relations in flux
compactifications, we find support for the possibility of a low
energy open string description of the static patch of de Sitter
space.
} \vskip .3in

\smallskip
\Date{April 2003}

\vfill \eject

\newsec{Introduction and Summary}

Flux compactifications play a very important role in string
theory. They provide examples of backgrounds with fixed moduli,
both in potentially realistic settings with small internal compact
spaces \refs{\KachruAW \GiddingsYU \FreyHF \KachruHE \MaloneyRR
\SilversteinXN - \AcharyaKV} and in canonical examples of the
AdS/CFT correspondence where the compact space is a large Einstein
space \MaldacenaRE. In both these cases, a negative contribution
to the effective moduli potential (coming from positive scalar
curvature, orientifold planes, or 7-branes wrapped nontrivially on
4-cycles in the base of elliptically fibred Calabi-Yau fourfold
compactifications of F theory \GiddingsYU) plays off against
positive contributions (including energy contained in quantized
fluxes) to produce a local minimum.  In a natural class of models
where we scale up the RR flux quantum numbers $Q_{RR}$ leaving
other parameters fixed, the string coupling is fixed at a value of
order
\eqn\basicgflux{ g_s\sim 1/Q_{RR} }

In this paper, we will study the entropy of flux
compactifications, focusing on the de Sitter case, via a simple
relation of de Sitter flux compactifications to deformations of
them which we will refer to as ``D-Sitter" spaces. These are
spacetimes containing D-brane domain walls surrounding a bubble of
a different de Sitter vacuum.

We will analyze explicitly basic properties of D-Sitter and
corresponding de Sitter flux compactifications, and apply this
analysis to obtain two basic results.

One of the main results will be a comparison of the entropy
carried locally on the D-branes, probed by an observer in the
middle of the bubble, with that of the original de Sitter space.
This will lead to an interesting string scale ``correspondence
point" at which they agree up to order one factors, somewhat
similar to the black hole correspondence point studied in
\refs{\SusskindWS,\HorowitzNW}.

The other main result will be circumstantial evidence pointing toward a low
energy open string theory in the de Sitter causal patch, with
of order $Q_{RR}$ ways for the open strings to end on the horizon.
This evidence will be twofold.  First, we will exhibit a set of
observers whose causal patch has a static coordinate system for
which one can take a limit in which the D-branes of the D-Sitter
space approach the horizon as the D-Sitter space approaches the
original de Sitter space. Second, we will show that a cutoff of
order string scale on $Q_{RR}^2$ D-brane worldvolume field theory
degrees of freedom, combined with the basic flux stabilization
result \basicgflux, produces an entropy agreeing with the de
Sitter entropy.  An open string picture of horizons has been
advocated in \SusskindSM, and we interpret our results as
providing some concrete but circumstantial
evidence for this picture, and an avenue toward
studying it explicitly to which we hope to return in future
work.\foot{Recent examples of other time dependent backgrounds
usefully described in terms of open strings have been proposed in
\SenIN\GutperleAI\ and many related works.}  Of course an open string
theory on the causal patch of
the perturbative de Sitter solution
does not account for decays
of de Sitter models \KachruAW\MaloneyRR\freyetal\ (which avoid
some of the puzzles
raised in e.g. \DysonNT\WittenKN\GoheerVF); we leave
for future work the description
of these decays from the D-Sitter point of view.

Having listed the main results, let us now turn to a
more extensive summary
of the motivation and the elements of our analysis.

In AdS/CFT examples, the physics of the flux compactification is
equivalent to that of a large $N$ dual field theory.  This
provides a holographic description of the background.  In
particular, the logarithm of the total number of states of the
cutoff field theory (which we will refer to as the Susskind-Witten
entropy) can be compared to that contained in AdS, and in the
simplest case of ${\cal N}=4$ super-Yang-Mills one finds an
entropy scaling like $N^2$ on both sides \SusskindDQ.

The relatively well understood AdS/CFT examples arise from near
horizon limits of brane systems, while generic flux
compactifications based on Calabi Yau or orbifold internal spaces,
even those that reduce to AdS as opposed to dS or Minkowski space,
have not been constructed from such a near horizon limit.  Also
the simplest AdS/CFT examples have a tunable dilaton, while
generically this modulus is fixed (or runs away) in flux
compactifications.

However, one can still study the entropy $S(\{ Q_i\} )$ associated
with a generic flux background as a function of the flux quantum
numbers $\{ Q_i \}$.  Moreover, one can trade fluxes for branes in
a region of the gravity background by introducing a bubble whose
wall is made up of D-branes.
%
\global\advance\figno by1
\ifig\flux{In a compactification of string theory with fluxes, we
can locally change the amount flux through a certain cycle by
introducing branes wrapped on the dual cycle.}
{\epsfxsize2.5in\epsfbox{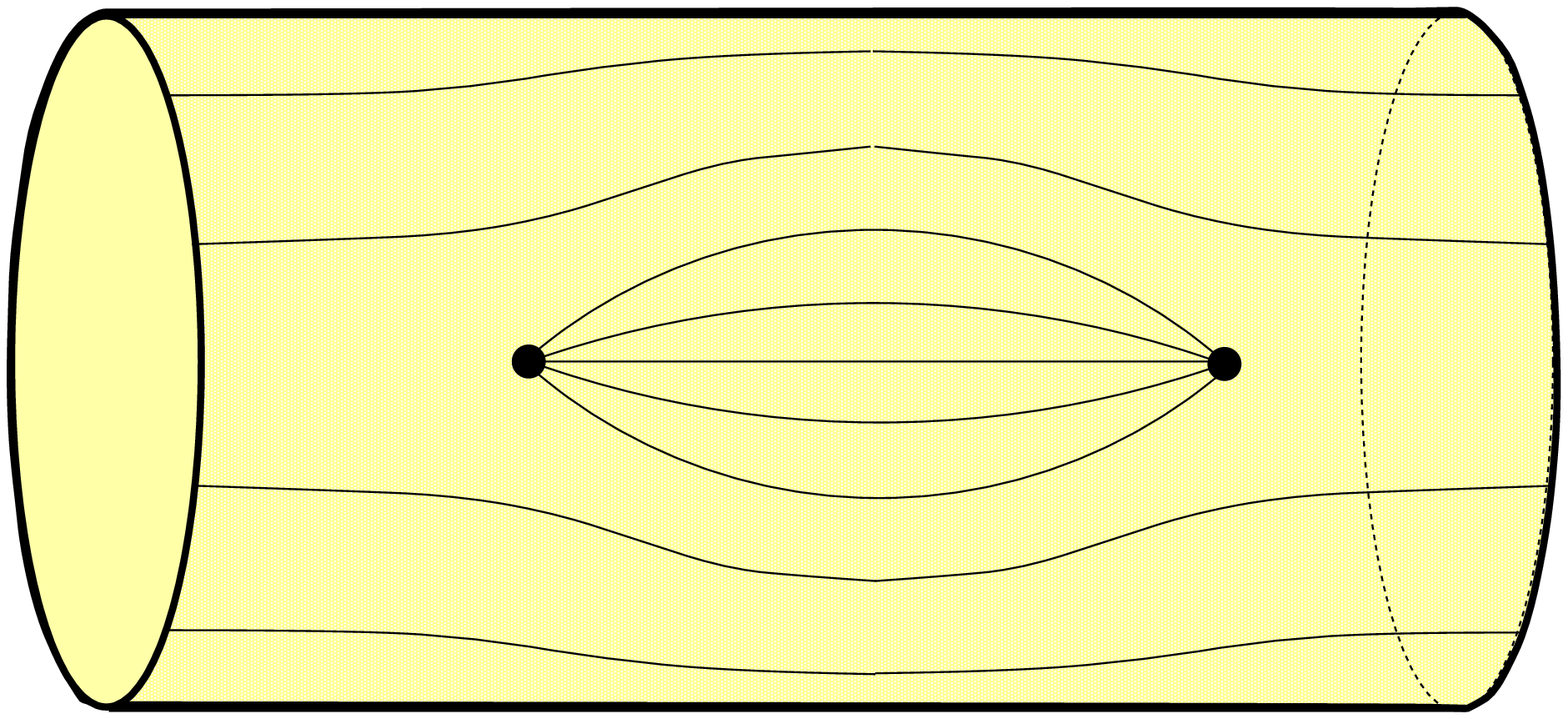}} In AdS/CFT examples, this
procedure produces the gravity dual of the field theory on its
Higgs or Coulomb branch, for which BPS states can be followed
adiabatically.  We can implement this procedure much more widely,
applying it to dS flux compactifications where we do not
know a field theory dual.

In the AdS/CFT case, this procedure could be applied step by step,
pulling a small number of D-branes out of the horizon in each
step.  Then calculating the spectrum of strings stretched to the
horizon from the branes would provide a microscopic accounting of
the derivative of the Susskind-Witten entropy with respect to flux
quantum numbers.  It is not yet known how to do this calculation
of the stretched string spectrum.  One can however pull of order
$N$ branes out of the horizon in the gravity dual of a large $N$
field theory (as in the simplest example of \KrausHV) and thus
transform of order $N^2$ of the entropy into states localized on
the D-branes.

In this paper, we will evaluate the entropy carried by D-brane
bubbles obtained by deformation from dS and AdS flux
compactifications, and compare it to the entropy associated to the
flux compactifications themselves.

To begin, we will analyze the thermodynamic and causal properties
of the de Sitter spacetimes containing bubbles (which we will
refer to as D-Sitter spacetimes), which turns out to be quite
interesting in its own right. In particular, a priori one might
think that there is no equilibrium thermodynamic ensemble in which
to compute the entropy because the branes separate phases of
different dS cosmological constant (and thus naively they feel
different temperature on the two sides of their worldvolume).
However, as we will see here, by taking into account the
acceleration of the brane observer on at least one side and the
resulting Rindler temperature, one can identify a well-defined
temperature on the D-branes and work consistently in a canonical
thermodynamic ensemble.

We will determine the causal patches of important classes of
observers in the D-Sitter spacetimes. One important type of
observer has a static causal patch identical to that of ordinary
de Sitter space, so that the deformation from de Sitter
to D-Sitter changes only
what is behind the horizon for this observer.
The observer in the middle of the bubble will
also play an important role.  This observer's causal patch is not
static but has very interesting properties: it contains the
D-branes which carry entropy, and at the same time has a smaller
horizon area at the moment of time symmetry than the original de
Sitter causal patch so that as we increase the brane entropy we
decrease the entropy associated with the horizon.

For generic
D-Sitter spacetimes, we will find that in the canonical ensemble
applying to our system,
the entropy localized on the
branes which is probed by this observer is much smaller than
the original de Sitter entropy.
Other excitations not localized on the D-branes,
such as strings stretching from the
D-branes to the horizon of this observers causal patch,
are Boltzmann suppressed in the canonical ensemble
but perhaps could play a role analogous
to that of the states counted in the full Susskind-Witten
entropy in the AdS case.

However, we will exhibit an interesting ``correspondence point"
related to that of \refs{\SusskindWS,\HorowitzNW} at which the
D-brane entropy approaches the dS entropy.  In this correspondence
limit, the Bousso bound \BoussoXY\ also approaches saturation for
all positive global times.



Brane observers have a static coordinate system parameterizing
their causal patch, as do observers maintaining a fixed distance
from the branes. For these latter observers, one can take a limit
in which the branes approach a trajectory tracking a portion of
the observer horizon and in which the observer's causal patch
approaches that of the original de Sitter space. This latter
result may provide a concrete avenue toward realizing the goal of
formulating horizon physics via open strings \SusskindSM.  Closed
strings exiting the horizon effectively have a boundary there, and
the relation to D-branes we uncover in this paper may help in
formulating this string theory as an open string theory.  As we
will see, the naive estimate that this putative open string theory
has of order $Q_{RR}$ Chan-Paton indices and should be described
as a field theory cut off at the string scale gives the correct
entropy for de Sitter space once we include the basic relation
\basicgflux\ coming from the flux stabilization of the dilaton in
real models.

This paper is organized as follows.  In section 2, we will explain
the deformation between flux compactifications and D-brane
bubbles, share a motivating analogy to our procedure in ordinary
AdS/CFT, and briefly discuss the Susskind-Witten entropy of more
general AdS flux vacua. In section 3, we turn to the de Sitter
case. We explain the causal structure and thermodynamics of the DS
(D-Sitter) spacetimes.  In section 4, we discuss the thermal
equilibrium of the domain walls. In section 5, we determine the
entropy carried by the branes as observed by probes sent from an
observer in the middle of the bubble, focusing on case of horizon
sized bubbles, and explain the ``correspondence point'' at which
this is comparable to the dS entropy.  In section 6, we
analyze Bousso's bound in D-Sitter spacetime.  In section 7, we
focus on the observers with a static causal patch and present our
circumstantial evidence for an open string interpretation.  In section 8 we
conclude with a summary and some discussion of open questions.

Other approaches to the microscopic counting of dS entropy have
appeared in \refs{\HawkingDA \MaldacenaIH \StromingerPN  \BanksFE
\HalyoPX \BalasubramanianRB \ParkQK - \ParkYW}. Other works have
recently studied nonperturbative decays of flux compactifications
arising when brane bubbles dynamically nucleate \refs{\FengIF
\BoussoXA \KachruGS - \freyetal, \MaloneyRR} following
\refs{\ColemanAW,\BrownKG} and many other works; a recent
discussion of thick wall decays occurs in \BanksNM.  In \Andreas\
appears an investigation of entropy associated with dS slices of
dS space.

\newsec{The basic deformation, and warmup AdS Examples}

One can rather generally obtain sets of D-branes associated to any
flux background by deforming the system to one containing D-brane
bubbles as in \flux. The procedure is the following. Let us
consider a compactification on $X$ down to $d$ dimensions. Given
flux $\int_C F=Q_R$ on a cycle $C$ of the compactification, we can
introduce $Q$ D-branes wrapped on a dual cycle $\tilde C$ sourcing
the same RR field strength $F$. This set of D-branes is locally a
domain wall in the $d$-dimensional spacetime, on one side of which
(say the right side) the flux is $Q_R$ and on the left side of
which the flux is $Q_L\equiv Q_R-Q$.  For these different flux
quantum numbers, the rest of the spacetime adjusts itself to solve
the equations of motion on each side in the presence of the
corresponding quantized flux.  Topologically this is always
possible, though generically this will lead to time dependent
solutions, and in some cases the evolution will take the system
out of theoretical control.

Our general strategy will be to consider cases where the resulting
D-brane system is under control, and to study the entropy of the
system both in the original flux compactification and in the
D-brane bubble spacetime obtained from this deformation.

In black hole physics, one could deform a set of D-branes into a
black hole by dialing the 't Hooft coupling $g_sQ$, and in
appropriate cases count the entropy precisely using the D-brane
worldvolume theory \StromingerSH.  While our goal (transmuting the
system into a system of D-branes whose states are easier to count)
is similar, note that the deformation we are considering is
different from that of \StromingerSH.  In order to deform from the
D-brane bubble spacetime into the original flux compactification,
we must shrink the bubble (which generically requires going over a
barrier).  As we will discuss in the next subsection, this is
analogous to moving on the moduli space of the field theory side
of AdS/CFT rather than dialing the 't Hooft coupling.

This difference with the black hole case of \StromingerSH,
combined with the basic difference that there is no unbroken
supersymmetry in our de Sitter case, will leave us with less
conclusive results than \StromingerSH.  Still, we will be able to
obtain results similar to the D-brane cases of the black hole
correspondence principle \refs{\SusskindWS,\HorowitzNW}, and we
will obtain results from flux compactification which are
nontrivially consistent with an open string interpretation of the
de Sitter causal patch.

D-brane domain walls have been applied fruitfully for example in
\refs{\GukovYA,\KachruGS,\AcharyaDZ} to describe important aspects
of the physics of gauge/gravity dual pairs, as well as to studying
nonperturbative decays of flux compactifications arising when
brane bubbles dynamically nucleate \refs{\FengIF \BoussoXA
\KachruGS - \freyetal, \MaloneyRR} following
\refs{\ColemanAW,\BrownKG}. Here, we will apply it to the question
of the microscopic accounting of the entropies associated with
flux compactifications.  We will focus on the de Sitter case in
later sections, but here we will begin with the very instructive
case of AdS.

\subsec{AdS/CFT analogy and motivation}

In the usual AdS/CFT correspondence, one can deform the gravity
side background smoothly into one in which there are D-branes
separating an AdS region from say flat space \KrausHV. In
the holographic dual field theory, this is a deformation
corresponding to going out on the Coulomb branch of the $U(N)$
${\cal N}=4$ super Yang-Mills theory. If we consider the entropy
corresponding to simply a count of the total number of states in
the Hilbert space below some cutoff \SusskindDQ, then we can
adiabatically follow the states as we move onto the Coulomb
branch.  In particular, for BPS states we expect of order $N^2$
massive off-diagonal BPS ``W boson'' states to be identifiable
after the deformation, and this corresponds to the fact that the
bubble wall of D-branes in the \KrausHV\ model carries entropy of
order $N^2$ from off diagonal stretched strings. Thus in this
case, the BPS states are still visible on the gravity side, but
have become massive stretched string states.\foot{In this example,
the non-BPS states are as always more difficult to follow, and
there appears to be substantial non-BPS entropy available in the
flat region inside the spherical shell of \KrausHV; e.g. one could
put a Schwarzschild black hole there.}  In other words, if we had
not known about the full AdS/CFT duality obtained by taking a near
horizon limit, but instead just had the $AdS\times S$ flux
compactification itself, we could still have obtained by this method a
nontrivial microscopic rendering of the BPS Susskind-Witten
entropy of the system in terms of degrees of freedom on D-branes.

\ifig\ads{Anti D-Sitter space:  a D-brane domain wall consisting
of $Q$ D3-branes separates
a region of flux $N_R$ to the right of the wall from a region of
flux $N_L=N_R-Q$ to the left of the wall. For $Q<< N_R$, the difference
in entropy at low energies goes like $2QN_L$, the number
of strings stretched from the branes to the horizon.  It is not
known how to calculate this number directly on the gravity
side of AdS/CFT.  For $Q$ of order $N_R$, on can count
states localized on the D-branes to obtain an entropy of the same
order in flux quantum numbers as the original anti de Sitter space.}
{\epsfxsize2.5in\epsfbox{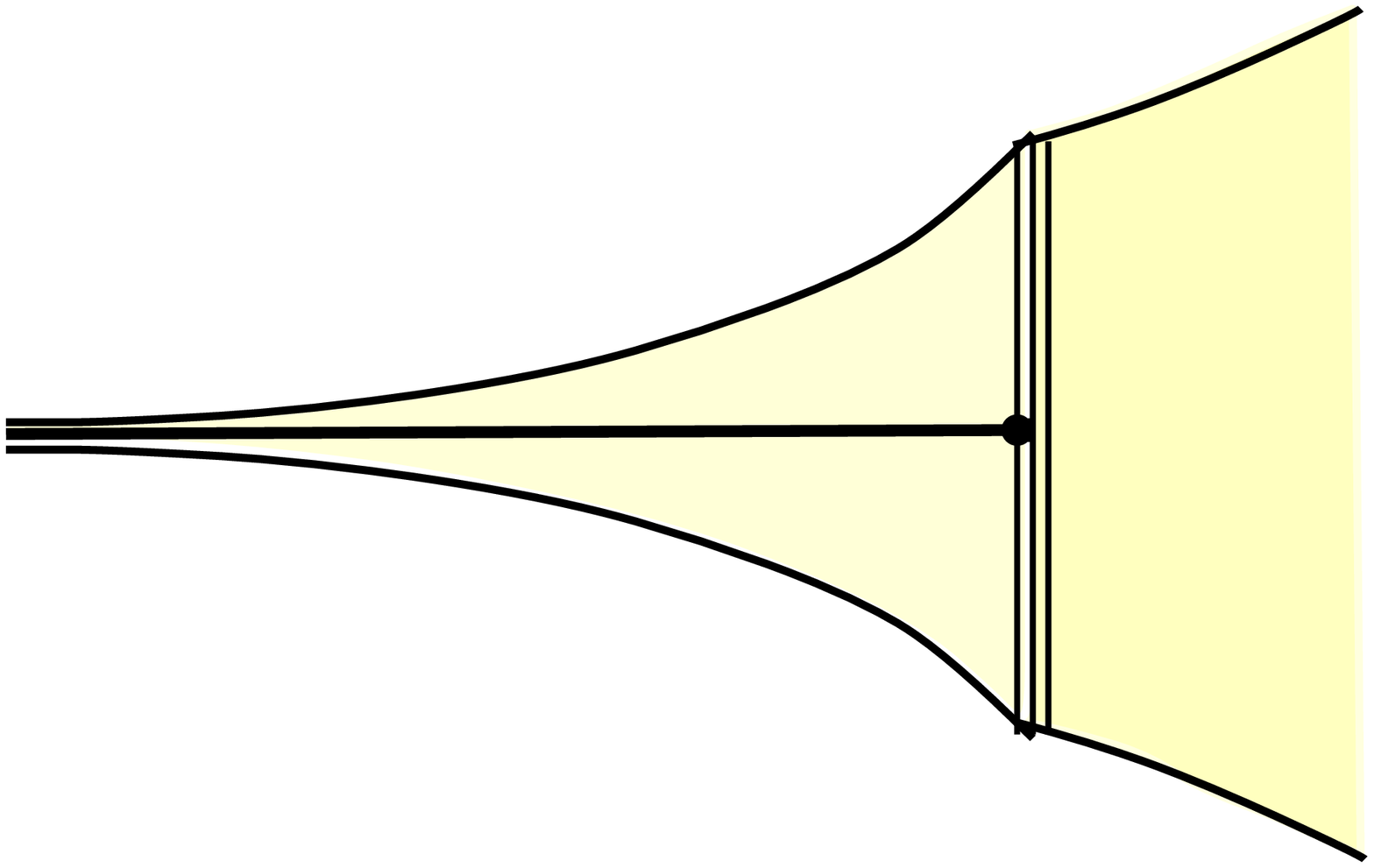}}

One can also study a Coulomb branch configuration corresponding to
taking $Q<<N_R$ branes out of an original stack of $N_R$ branes
and separating them from $N_L=N_R-Q$ remaining branes. This
corresponds in the near horizon region to a domain wall containing
$N$ D3-branes separating a phase of flux $N_R$ from one of flux
$N_L\equiv N_R-Q$.  The change in the number of degrees of freedom
as we pull the $Q$ branes away is of order $Q N_L$ and comes
mostly from strings stretching from the domain wall to the horizon
of the Poincare patch.  If we could independently count these
strings purely on the gravity side, then this would provide a
microscopic accounting of the derivative of the entropy with
respect to the flux quantum number.  It is not known how to do
this counting; in any case in a fixed temperature ensemble these
strings are Boltzmann suppressed. As we have seen, in the AdS case
we can simply trade all of the flux for branes in the flat region
of the \KrausHV\ examples.

In our de Sitter case, we will also not be able to count such
stretched strings and will instead focus on the question of how
many states are localized on the D-brane wall itself. These
strings alone will be able to saturate the dS entropy in a string
scale ``correspondence point".  It would be very interesting to
develop techniques to count the strings stretched to the horizon
in both the AdS and dS cases.  We will leave this for future work.


An interesting variant on this case is to carry out the same
procedure for situations with fractional branes, i.e. orbifolds of
AdS/CFT \KachruYS\ such as $AdS_5\times S^5/Z_k$. Then although
the flux integrated over the cycle $S^5/Z_k$ is $N$, the total
entropy is $kN^2$ rather than $N^2$ as our naive estimate would
indicate. In this case, one could deduce this from the D-brane
domain wall configuration in the near horizon limit by the
presence of $k$ sectors of wound strings.

AdS vacua having no known interpretation as near horizon limits of
branes arise from flux compactifications as discussed recently in
various approaches \refs{ \KachruAW, \AcharyaKV}. In the KKLT
models yielding AdS vacua (i.e. before the introduction of the
anti D3-brane which uplifts the AdS to dS in their construction),
the Susskind-Witten entropy as a function of flux quantum numbers
can be estimated. As we will show in \S5.4, the Bousso-Polchinski
mechanism allows one to tune the cosmological constant very
finely, leading to an entropy that can be as large as of order
\eqn\entKKLT{ S\sim Q^{{\chi\over 2}+4} }
in terms of a flux quantum number $Q$ where $\chi$ is the Euler
characteristic of the Calabi-Yau fourfold compactification of F
theory. The result \entKKLT\ is much larger than the naive entropy
on D-branes of order $Q^2$. However, it is intriguing in its
dependence on the integer flux quantum numbers, perhaps suggesting
a dual gauge theory with many flavors. It is conceivable that this
indicates a phenomenon like that of fractional branes just
reviewed.




\newsec{Causal Structure and Penrose Diagrams}

Let us now move on to our main interest of de Sitter flux vacua.
We will be interested in the entropy carried by the D-branes in
the D-Sitter spacetimes which are deformations of de Sitter space
with D-brane bubbles. In ordinary de Sitter space, each observer
determines a causal patch, and the horizon of area $A$ in this
causal patch has been argued to carry an entropy accessible to
this observer $A/4l_d^{d-2}$ in terms of the $d$-dimensional
Planck scale $l_d$ \GibbonsMU.  As we discussed in \S1, we can
deform dS to DS by introducing D-brane bubbles.  In this section
we will begin our analysis of these spacetimes by determining
their causal structure, including an explicit specification of the
Penrose diagrams for D-Sitter.  This will enable us to determine
the causal patches for various observers which replace the causal
patch of the original de Sitter space for those observers. For an
important class of D-Sitter solutions, there is at least one
observer whose causal patch remains the same as in the original de
Sitter space.  For an observer in the middle of the bubble, the
causal patch has a smaller horizon area than the original de
Sitter causal patch, and in the case of a bubble of flat space the
size of the original dS horizon this area shrinks to zero.  For
observers maintaining a fixed distance from the branes, there is a
static coordinate system covering their causal patches.  This
causal patch has the interesting property that one can take a
limit in which the D-branes approach the horizon for these
observers.

\subsec{Penrose Diagrams}

\ifig\instanton{This figure shows a solution of the Euclidean
Einstein's equations, with two regions of different cosmological
constants ($\Lambda_L < \Lambda_R$) separated by a massive brane.
In the thin wall approximation, matching the metrics on two sides
of the brane is done using the Israel junction conditions. By
Wick-rotating the $\theta$ coordinate one obtains a solution of
Lorentzian Einstein's equations which contains two different de
Sitter vacua connected to each other by a brane. We refer to such
Lorentzian solutions as `D-Sitter'. }
{\epsfxsize2.2in\epsfbox{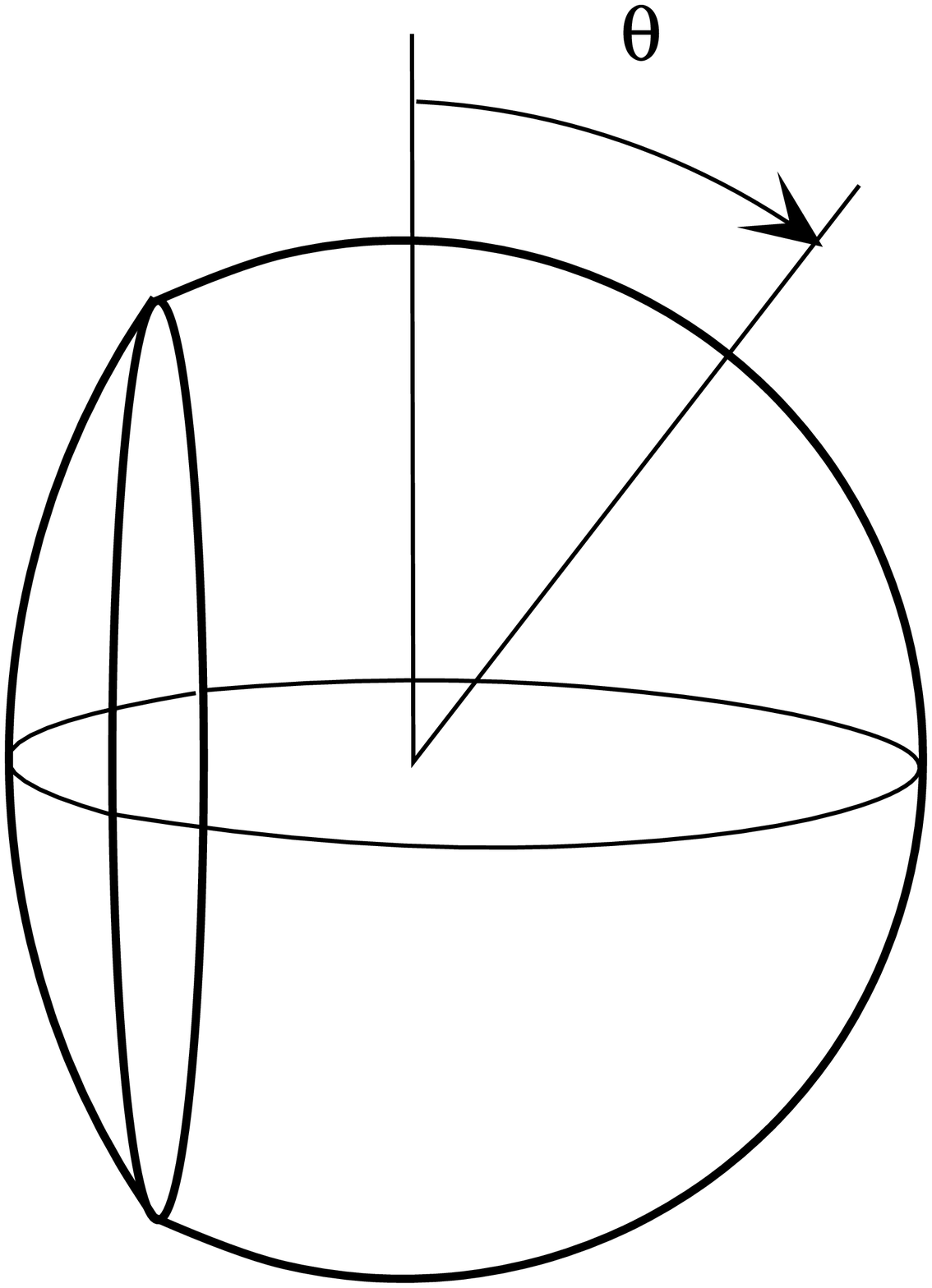}} 

Let us now analyze the causal structure of the D-Sitter spacetimes
which arise as analytic continuations of the Euclidean solutions
from \instanton to Lorentzian signature. These spacetimes contain
a spherical domain wall which first contracts, and then  expands
again. The domain wall separates two regions (referred to as
$dS_L$ and $dS_R$) of different cosmological constants
($\Lambda_L$ and $\Lambda_R$). Our strategy will be to
analytically continue $\theta$ in each of those regions
separately. We will work mainly in the thin wall approximation,
and comment on the thick wall generalization at the end.


Let us consider one particular side of the Euclidean solution. The
round sphere metric is given by
\eqn\sphereone{ds^2 = d\theta^2 + \sin^2\theta \ d\Omega^2_{d-1}.}
After the analytic continuation $\theta \to i\tau + \pi/2$, this
becomes de Sitter space in global coordinates
\eqn\globaldesitter{ds^2 = -d\tau^2 + \cosh^2\tau \
d\Omega^2_{d-1}.}
It is convenient to define a new time coordinate -- the conformal
time $T$
\eqn\conformaltime{{1 \over \cos T} =  \cosh \tau, \quad T \in
\left(-{\pi \over 2},{ \pi \over 2}\right),}
and to expand $d\Omega^2_{d-1}$ as
\eqn\globaldesittertwo{ds^2  = {1 \over \cos^2 T }\ \! \left(- \
\! dT^2 + d\thp^2 + \sin^2 \thp \ \! d\Omega^2_{d-2} \right).}
\ifig\desitter{Penrose diagram of de Sitter space. Over each point
in this picture there is an $S^{d-2}$, which degenerates to a zero
size at $\theta' = 0$ and $\theta' = \pi$. Every horizontal slice
is an $S^{d-1}$. The diagonal lines represent light rays
originating from the poles of the $S^{d-1}$ at $T = - \pi /2$. }
{\epsfxsize2.2in\epsfbox{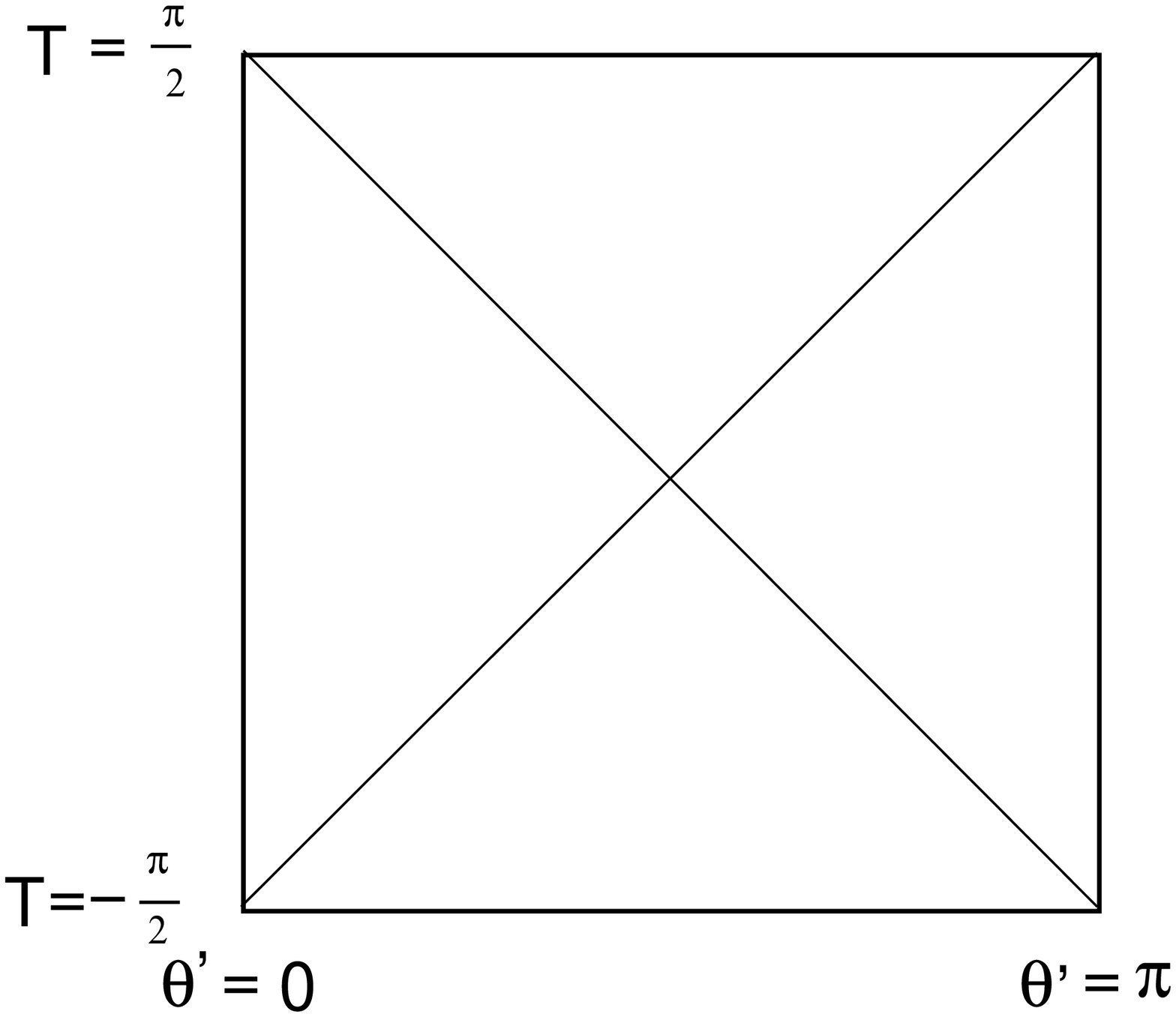}}
\noindent Now, suppressing the $S^{d-2}$ directions, we obtain the
Penrose diagram in \desitter. Only a part of it will be relevant
for us at this point.

In a full global de Sitter, the range of the coordinate $\thp$
would be $(0, \pi)$. In the D-Sitter spacetimes it will be
restricted to
\eqn\thetaprimerestriction{\thpl \in (0,\ \! \thplb(T_L)), \quad
\thpr \in ( \thprb(T_R),\ \! \pi)}
 on the left and  on the right, respectively, because of the presence
 of the domain wall. We would like to find the precise form of the
time-dependence of the position of the domain wall (or the
`bubble'). In other words, we want to identify the functions
$\thplb(T_L)$ and $\thprb(T_R)$.

This can be easily done using the original Euclidean solution. Its
embedding $(d+1)$-dimensional flat space may be expressed as
\eqn\embeddingone{\eqalign{x_1 &= R \cos\theta \cr x_2 &=
R\sin\theta\cos{\theta'}  \cr x_\alpha &= R \sin\theta \sin\theta'
n_\alpha, \quad \alpha = 3 \dots (d+1), }  }
where  $\sum_\alpha  n_\alpha^2 = 1$. The coordinates $x_i$
satisfy
\eqn\spheretwo{x_1^2 + x_2^2 + \sum_{\alpha = 3}^{d+1} x_\alpha^2
= R^2.}
The worldvolume of the domain wall is the intersection of
\spheretwo\ with
\eqn\planarcut{x_2 = { \pm \sqrt{R^2 - R_B^2}},}
where only one sign on the right hand side should be considered.
From \embeddingone\ and \planarcut\ we obtain
\eqn\costhetaprimeone{\cos \theta' = \pm \sqrt{1-{R_B^2 \over
R^2}} \ \ {1 \over \sin \theta} .}
After the analytic continuation $\theta \to i\tau + \pi/2$ this
becomes (using \conformaltime)
\eqn\costhetaprimetwo{\cos \theta' =
 \pm \sqrt{1-{R_B^2 \over R^2}} \
\  \cos T.}
%
%
%
%
%
%
Thus we have obtained the explicit time-dependence of the position
of the domain wall
\eqn\bubbletrajectoryleft{\cos \thplb (T_L) = \pm \sqrt{  1-{R_B^2
\over R^2_L }  } \ \! \cos T_L,}
\eqn\bubbletrajectoryright{\cos \thprb (T_R) = \pm \sqrt{ 1-{R_B^2
\over R^2_R }  } \ \! \cos T_R.}
All combinations of signs are possible here except having a minus
sign in \bubbletrajectoryleft\ and a plus sign in
\bubbletrajectoryright. We will always choose $R_L > R_R$.
\ifig\bubbledesitter{D-Sitter spacetimes with a (a) `subcritical'
bubble, $\thprb(T_R) < \pi /2$, (b)  `critical' bubble,
$\thprb(T_R) = \pi /2$, and (c)  `supercritical' bubble
$\thprb(T_R)
> \pi /2$. In all cases we choose $\Lambda_L < \Lambda_R$, which means $R_L >
R_R$. The shaded regions are absent from the spacetimes. In each
case the two boundary lines of the shaded regions are to be
identified.}
{\epsfxsize5in\epsfbox{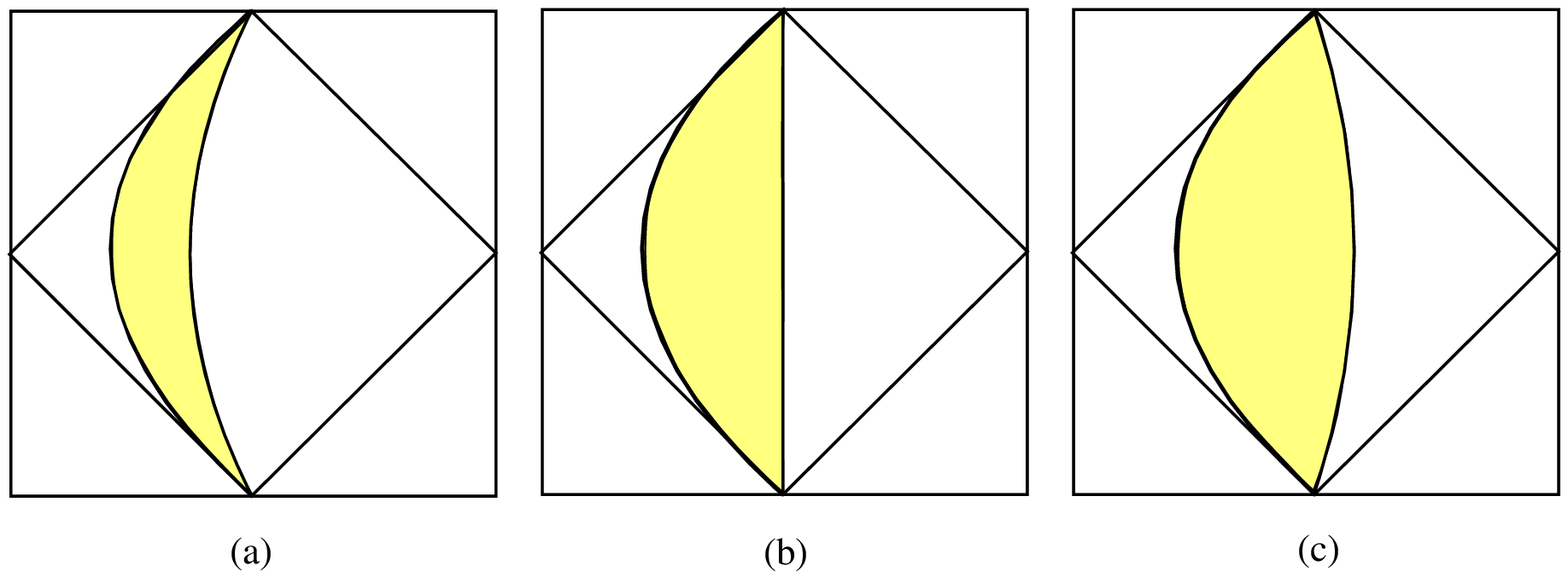}} 
\noindent The corresponding Penrose diagrams are depicted in
\bubbledesitter.

Having determined the position of the domain wall for each side in
terms of $T_L$ and $T_R$ respectively, we can now match the two
solutions by determining how the left and right coordinates on the
bubble should be identified.  That is, we would like to know which
time $T_L$ at the domain wall corresponds to which $T_R$ at the
same point. Again, this can be easily found using the original
Euclidean solution where (cf. \embeddingone)
\eqn\matchingeuclideantimes{R_L \cos \theta_L = R_R \cos
\theta_R.}
After the analytic continuation we obtain
\eqn\machingtaus{R_L \ \! \sinh \ \! \tau_L = R_R \ \! \sinh \ \!
\tau_R.}
Expressed in terms of time $T_L$ and $T_R$ at the bubble this is
\eqn\matchingtimes{{1\over \cos^2 T_L}=1 + {R_R^2 \over R_L^2}
\left( {1 \over \cos^2 T_R} - 1 \right).}

These relations are general (given the thin wall approximation).
It is useful to distinguish three types of bubbles depending on
whether the bubble is at the horizon size at $\tau=0$, below the
horizon size, or above. Let us take for simplicity the case that
$\Lambda_L \ll \Lambda_R$. (As we will review in \S5, it is
possible in flux models to discretely tune $\Lambda$ to be very
close to zero \refs{\BoussoXA,\MaloneyRR,\KachruAW}.) This means
that the bubble is superhorizon size from the point of view of the
$dS_L$.  From the point of view of the $dS_R$, the bubble can be
subhorizon sized, horizon sized, or superhorizon sized.  As
reviewed in \MaloneyRR, these cases correspond to a brane tension
$T_B$ whose square is less than, equal to, or greater than a
critical value
$T_c^2={2(d-2)(d-1)^{-1}(\Lambda_R-\Lambda_L){l_d^{4-2d}}}$
\BrownKG. We will refer to these bubbles as subcritical, critical,
and supercritical, respectively.

The Penrose diagrams we have determined so far (\bubbledesitter)
had two separate pieces, each corresponding to one side of the
domain wall, along with a prescription for identifying the points
on the bubble wall to join the two pieces together. This might be
sufficient for our purposes since we know how to map the points of
the domain wall from one part of the Penrose diagram to the other.
However, it may still be interesting to see how to construct one
global (and connected) Penrose diagram from the two parts. In
order to do so, we will have to perform a (conformal) coordinate
transformation at least on one side.

Our strategy will be the following. We want to find a global
coordinate system $(\thpg, T_G)$ such that in the left part of the
spacetime $(\thpg, T_G)$ are functions of $(\thpl, T_L)$, and in
the right part of the spacetime $(\thpg, T_G) = (\thpr, T_R)$. We
require that also in the new $(\thpg, T_G)$ coordinate system the
light-cones  are  at 45 degrees. This means, of course, that the
transformation from $(\thpl, T_L)$ to $(\thpg, T_G)$ must be
conformal, i.e. of the form
\eqn\conformaltransformation{T_G + \thpg = f(T_L + \thpl), \quad
T_G - \thpg = g(T_L - \thpl), }
where $f$ and $g$ are some functions. We have to make sure that at
the domain wall the coordinates match correctly onto the right
part of the Penrose diagram, where we decided to keep the original
coordinates (i.e. where we chose $(\thpg, T_G) = (\thpr, T_R)$).
This requirement translates into
\eqn\matchingone{T_R + \thprb (T_R) = f(T_L + \thplb (T_L), ) }
and
\eqn\matchingtwo{T_R - \thprb (T_R) = g(T_L - \thplb (T_L) ) ,}
where the functions $\thplb (T_L)$ and $\thprb (T_R)$ are given by
\bubbletrajectoryleft\ and \bubbletrajectoryright, and where the
times $T_L$ and $T_R$ are related by \matchingtimes. The
conditions \matchingone\ and \matchingtwo\ determine the explicit
form of the functions $f$ and $g$ in the conformal transformation
\conformaltransformation.
\ifig\connected{Another rendering of the Penrose diagrams of
D-Sitter spacetimes with a (a) `subcritical' bubble, $\thprb(T_R)
< \pi /2$, (b) `critical' bubble, $\thprb(T_R) = \pi /2$, and (c)
`supercritical' bubble $\thprb(T_R)
> \pi /2$. In all cases we choose $\Lambda_L < \Lambda_R$, which means $R_L >
R_R$.} {\epsfxsize5in\epsfbox{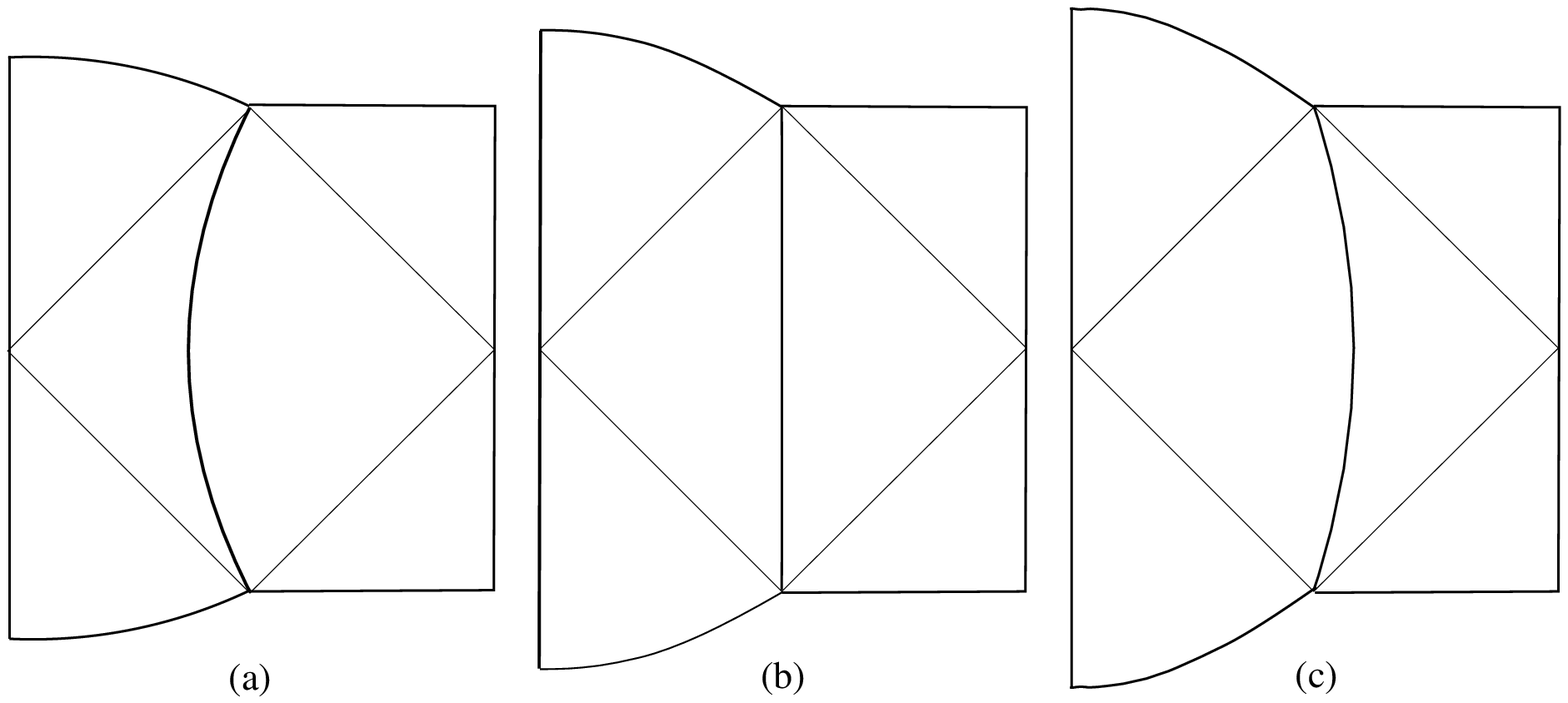}} 
\noindent The resulting connected Penrose diagrams are in
\connected.

So far we have considered the idealized case where the domain wall
is infinitely thin. One can also construct Penrose diagrams for
spacetimes where the domain wall has a finite thickness. The
causal structure is nevertheless very similar to the idealized
case. (This case has been analyzed recently in \BanksNM\ following
\ColemanAW\ in the context of bubble nucleation.)
The analytic continuation of the Euclidean solution
for a thick wall would suggest an exponentially growing
wall thickness in the Lorentzian continuation.  However,
as long as the force holding the brane together is stronger
then that coming from the de Sitter expansion, in
the stable Lorentzian solution the brane's
thickness will not grow exponentially and its intersection
with future infinity in the Penrose diagram will be just
a point.

\subsec{Causal Patches in D-Sitter}

We now turn to the question of the causal patches for observers in
the D-Sitter spacetimes.  In the Penrose diagram for ordinary de
Sitter space (\desitter) each point represents an $S^{d-2}$, which
shrinks to zero size at the left and right edges of the diagram.
These edges are natural geodesics on which to place observers
(both in $dS$ and $DS$)  which we will refer to as ``left" and
``right" observers ${\cal O}_L$ and ${\cal O}_R$ respectively. In
$dS$, each of these observers determines a causal patch, the
triangles on the left and right sides of the figure. The static
patch coordinates
\eqn\statcoords{ ds^2=-\left(1-{r^2\over
R_{dS}^2}\right)dt^2+{dr^2\over 1-{r^2\over
R_{dS}^2}}+r^2d\Omega^2_{d-2} }
reveal a horizon at $r=R_{dS}$.  The time coordinate $t$ in the
static patch goes to $\pm\infty$ at the two interior edges of the
triangle (at $T=\pm(\theta'-\pi/2$)), and goes through zero at the
central vertex of the triangle (at $T=0,\theta'=\pi/2$).  At
$t=0$, $r=1$ the transverse $S^{d-2}$ has area $A\sim
R_{dS}^{d-2}$. Here sit the states accounting for the de Sitter
entropy according to the ``hot tin can" picture developed e.g. in
\DysonNT\ (which we will briefly review in \S6).

We would like to know what the analogous results are for the
D-Sitter spacetime.  In the D-Sitter spacetime, there are several
interesting classes of observers.

\bigskip
\noindent{\it Left Observer's Causal Patch}
\bigskip

Let us first focus on the observer ${\cal O}_L$ in the center of
the bubble in the $dS_L$ part of the space.  For subcritical
bubbles and critical bubbles, the right observer ${\cal O}_R$
has the same causal patch as in ordinary $dS_R$.

In particular, we would like to compute the geometry of the causal
patch and the horizon area of ${\cal O}_L$
\ifig\horizon{Calculation of the horizon area of the observer
${\cal O}_L$
sitting on the left of the Penrose diagram, i.e. at $\theta' = 0$.
} {\epsfxsize2.2in\epsfbox{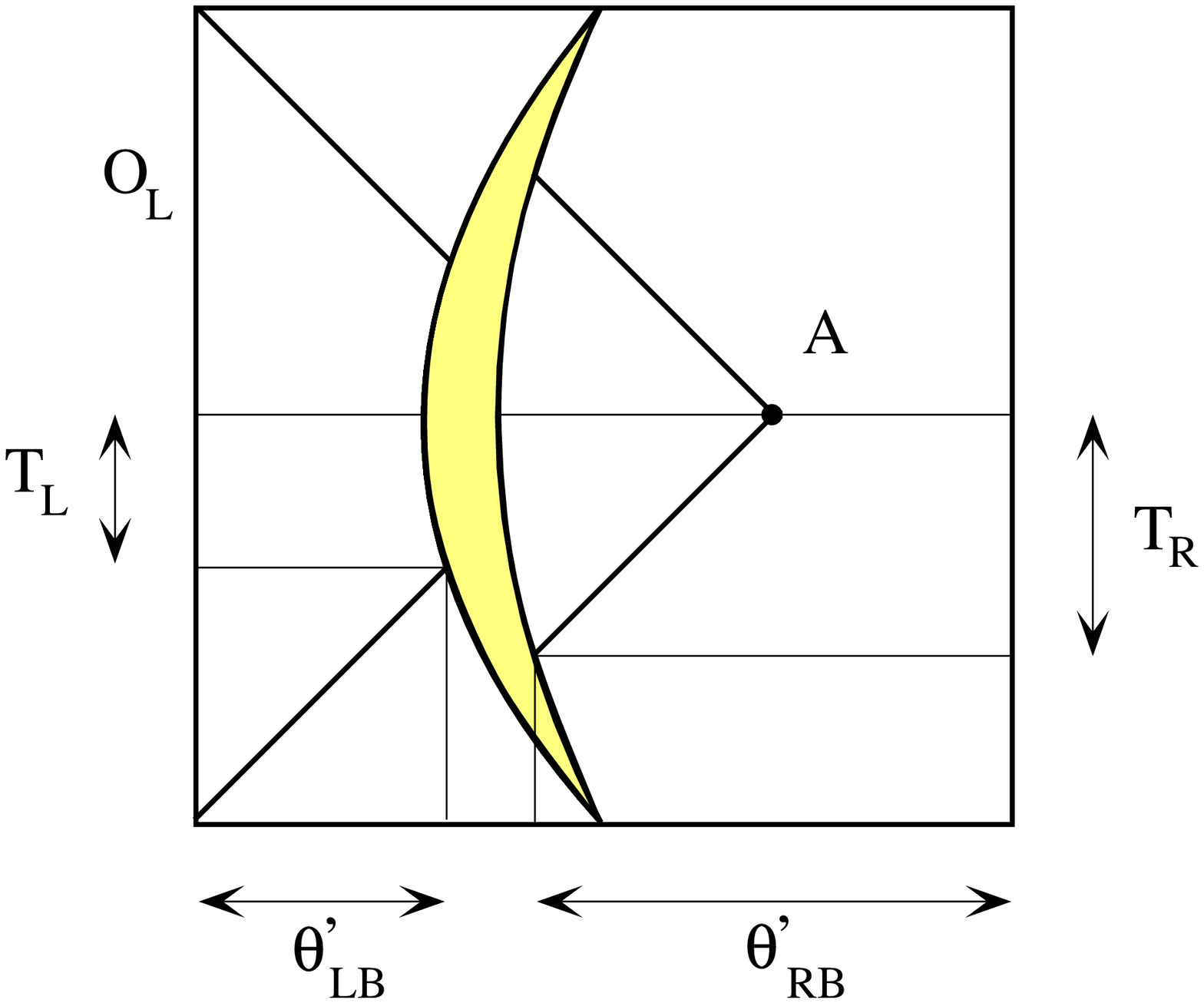}}
(see \horizon).

First, we need to find the time $T_L$ when the light ray $\CL_1$
reaches the domain wall,
\eqn\hitwallone{\thplb (T_L) = T_L + {\pi \over 2} .}
Using \bubbletrajectoryleft\ we can rewrite \hitwallone\ as
\eqn\hitwalltwo{{1\over \cos^2 T_L} = \left( 2 - {R_B^2 \over
R_L^2}  \right).}
This value of $T_L$ translates via \matchingtimes\ into the
following $T_R$,
\eqn\hitwallthree{{1\over \cos^2 T_R} = {R_L^2 + R_R^2 - R_B^2
\over R_R^2},}
and from \bubbletrajectoryright\ we see that the corresponding
$\thprb$ satisfies
\eqn\hitwallfour{\sin^2 \thprb  = {R_L^2 \over R_L^2 + R_R^2 -
R_B^2 }.}
A short examination of \horizon\ reveals that the horizon are of
the left observer at $T=0$ is given by
\eqn\horizonarea{A= |\Omega_{d-2}| R_R^{d-2} \sin^{d-2} (\thprb +
T_R ),}
 where $T_R$ and $\thprb$  are given by \hitwallthree\ and
 \hitwallfour, and where $|\Omega_{d-2}|$ stands for the volume of
 a unit $S^{d-2}$.

Notice that for $R_L>>R_R$, $sin^2\theta'_{RB}\to 1$ so
$\theta'_{RB}\to\pi/2$ while $cos^2T_R\to 0$ so $T_R\to -\pi/2$.
This means $A\to 0$.  In general, the horizon area we have
calculated satisfies $A<A_R$, so adding branes (which carry some
entropy) goes along with a decrease in the horizon area.  We will
exhibit an explicit relation expressing this tradeoff in \S5.1.

We can simplify this result to obtain
\eqn\simpleA{
A=|\Omega_{d-2}|R_R^{d-2}\left({R_LR_R+\sqrt{(R_R^2-R_B^2)(R_L^2-R_B^2)}\over
R_L^2+R_R^2-R_B^2}\right)^{d-2} }
For critical bubbles, $R_R=R_B$ and
$A=|\Omega_{d-2}|R_R^{d-2}(R_R/R_L)^{d-2}$.

Note that the causal patch we have derived for the left observer
${\cal O}_L$ is not a {\it static} patch: the proper area of the
bubble wall grows in time and so one cannot find static
coordinates describing the whole left causal patch containing the
bubble. The horizon area also changes as a function of time in the
left causal patch.  Both the D-brane entropy and the entropy
associated with the horizon area grow in time for $\tau>0$.

There is a set of observers whose causal patch is also static;
these are the brane observers and those at a fixed distance from
them, to which we turn next.

\bigskip
\noindent{\it Static patches in D-Sitter}
\bigskip



In  ordinary de Sitter space, one can identify a certain part of
the space -- referred to as the `static patch' -- in which there
exists a time-like Killing vector field. One can find many
distinct static patches in a given global de Sitter space (cf.
figs. 11 and 12). All of them are however equivalent up to the
action of the de Sitter group $SO(d-1,1)$. Each static patch turns
out to be the causal patch of some observer ( i.e. it is the set
of points which are in both the causal past of at least one point
of the observer's worldline, and in the causal future of some
other point of the observer's worldline).

We would like to find the static patches in D-Sitter spacetimes.\foot{We
would like to thank L. Susskind for a question leading us
in this direction.}
Of course, in the case of the subcritical bubble (\bubbledesitter (a)),
the causal patch of the right observer will be static,
and it will not contain the bubble at all. Depending on the
precise shape of the bubble, this will also be  true for other
observers with $\theta'$ not too far from $\pi$. There can be,
however, more static patches in the D-Sitter space.

The static patch of the usual de Sitter space (and its natural
coordinate system) can be obtained by an analytic continuation
from the Euclidean de Sitter solution, i.e. from a $d$-dimensional
sphere. One writes the $S^d$ as an $S^1$ fibration over a
hemisphere $S^{d-1} / \IZ_2$ in such a way that there is a
manifest $U(1)$ isometry corresponding to motions along the $S^1$
fiber. The metric on the $S^{d-1} / \IZ_2$ base is chosen to be
independent of the fiber coordinate $\phi_d$. Then by
Wick-rotating $\phi_d$ one obtains a patch of de Sitter space
which is static. The static patch extends up to a horizon, located
at the boundary of the $S^{d-1} / \IZ_2$ (where in the Euclidean
solution the $S^1$ fiber was degenerate).

We will repeat this construction here. Eventually, we will
restrict our coordinates to range only up to the place where we
want to place the domain wall. In this way we will obtain one side
of a static patch in D-Sitter.

Let us choose the following coordinates for the $S^d$.
\eqn\spherethree{\eqalign{y_1 &= R \cos \phi_1 \cr y_2 &= R \sin
\phi_1 \cos \phi_2 \cr y_3 &= R \sin \phi_1 \sin \phi_2 \cos\phi_3
\cr &\dots \cr y_d &= R \sin \phi_1 \sin \phi_2 \sin \phi_3 \dots
\cos \phi_{d} \cr y_{d+1} &= R \sin \phi_1 \sin \phi_2 \sin \phi_3
\dots \sin \phi_{d} }}
The $\phi_1, \dots \phi_{d-1}$ parameterize the $S^{d-1} / \IZ_2$
base, and the $\phi_d$ is the $S^1$ fiber coordinate.
 The range of
$\phi_i$ for $i < d$ is $[0, \pi)$, and the range of $\phi_d$ is
$[0,2\pi)$.
The metric of the $S^d$ can be written as
\eqn\spheremetric{\eqalign{{ds^2\over R^2}  = d\phi_1^2 &+ \sin^2
\phi_1 \ \! d\phi_2^2 + \sin^2 \phi_1 \sin^2 \phi_2 \ \! d\phi_3^2
+ \dots \cr &+ \sin^2 \phi_1 \sin^2 \phi_2 \sin^2 \phi_3 \dots
\sin^2 \phi_{d-1} \ \! d\phi_d^2. }}
By the Wick rotation $\phi_d \to i\tilde t$, we get the metric of
the de Sitter static patch
\eqn\staticmetric{\eqalign{{ds^2\over R^2} = d\phi_1^2 &+ \sin^2
\phi_1 \ \! d\phi_2^2 + \sin^2 \phi_1 \sin^2 \phi_2 \ \! d\phi_3^2
+ \dots \cr &- \sin^2 \phi_1 \sin^2 \phi_2 \sin^2 \phi_3 \dots
\sin^2 \phi_{d-1} \ \! d\tilde t^2. }}

Now we will consider the spatial geometry at some definite $t={\rm
const. }$ The horizon will be located at
\eqn\horizonpositionone{ \sin^2 \phi_1 \sin^2 \phi_2 \sin^2 \phi_3
\dots \sin^2 \phi_{d-1}=0.}
Even though this condition looks complicated, it is actually
equivalent to
\eqn\horizonpositiontwo{ \sin \phi_{d-1}=0,}
because if $\sin \phi_{i} = 0$ for any $i<{d-1}$, the $\phi_{d-1}$
coordinate becomes degenerate, and we may as well say that $\sin
\phi_{d-1}=0$. Given the fact that we chose $\phi_{d-1}$ to be in
$[0,\pi)$, we can rewrite \horizonpositiontwo\ as
\eqn\horizonpositiontwo{  \phi_{d-1}={0}
.}
In terms of the embedding coordinates of the $S^{d-1} / \IZ_2$
\eqn\spatialsphere{\eqalign{\tilde y_1 &= R \cos \phi_1 \cr \tilde
y_2 &= R \sin \phi_1 \cos \phi_2 \cr \tilde y_3 &= R \sin \phi_1
\sin \phi_2 \cos\phi_3 \cr &\dots \cr \tilde y_{d-1} &= R \sin
\phi_1 \sin \phi_2 \sin \phi_3 \dots \cos \phi_{d-1} \cr \tilde
y_{d} &= R \sin \phi_1 \sin \phi_2 \sin \phi_3 \dots \sin
\phi_{d-1} }}
(which have to satisfy $\tilde x_{d}\ge 0$) the condition
\horizonpositiontwo\ is simply
\eqn\horizonpositionthree{\tilde y_{d} = 0.}
For $d-1=2$ we could say that the horizon (i.e. the $\tilde y_{d}
= 0$ curve) is well approximated by the union of the Greenwich
meridian and the international date line (see fig. 8). For an
observer living in the US, the world is terminated there, and the
eastern hemisphere does not exist. In higher dimensions, the
$\tilde y_{d} = 0$ curve  becomes an $S^{d-2}$ surface dividing
the $S^{d-1}$ into two halves. Only one half has a  physical
significance.

For the reader's convenience, we will also provide an explicit
coordinate redefinition which transforms the metric in the de
Sitter static patch \staticmetric\ into the most standard form.
Notice that using \spatialsphere\ we can rewrite \staticmetric\ as
\eqn\staticmetricy{ds^2 = - \tilde y^2_d \ \! d\tilde t^2 +
\sum_{i=1}^d d \tilde y_i^2, \quad \sum_{i=1}^d \tilde y_i^2 =
R^2,}
with the indicated constraint imposed on the  $\tilde y_i$
coordinates. It is easy to check that with the following
definitions
\eqn\redefinition{ t = R\tilde t, \quad r^2 = R^2 - \tilde y_d^2,
\quad \Omega_i = {\tilde y_i \over \sqrt{R^2 - \tilde y_d^2}},
\quad i = 1, \dots (d-1),}
the static patch metric \staticmetric\ becomes
\eqn\standardmetric{ds^2 = - \left( 1- {r^2 \over R^2} \right)
dt^2 + {dr^2 \over {1 - r^2/R^2} }+ r^2 d\Omega^2.}

\bigskip

Now we can ask where we can place the domain wall inside the
static patch. In the Euclidean solution, the brane could have been
at the intersection of the sphere \spherethree\ with any plane
\eqn\plane{a_1 y_1 + a_2 y_2 + \dots + a_d y_d + a_{d+1}y_{d+1} =
{\rm const.}, }
such that the radius of the spherical cut is $R_B$. However, we
have Wick-rotated $\phi_d$. Because the definition of $y_d$ and
$y_{d+1}$ contains $\phi_d$, we must set $a_d = a_{d+1} = 0$ in
\plane. Otherwise, the brane would not be static in the `static
patch'. Thus we can express the position of the brane purely in
terms of the $\tilde y_i$ ($i<d$) coordinates defined in
\spatialsphere,
\eqn\planetwo{a_1 \tilde y_1 + a_2 \tilde y_2 + \dots + a_{d-1}
\tilde y_{d-1} = {\rm const.} }
We  can also  use an $SO(d-1)$ coordinate redefinition to
transform \planetwo\ into
\eqn\planethree{ \tilde y_{d-1}  = {\rm const.} }
\ifig\hemispheres{The shaded surfaces in this figure represent the
spatial geometry on one side of the brane in the static patch at
some definite $t={\rm const.}$ To obtain the full spatial geometry
of the static patch, one needs to glue together two surfaces of
this type. The different cases in this figure are chosen to match
the three possibilities for the right part of the diagrams in
\figs{\bubbledesitter\ and\ \connected}.
} {\epsfxsize5in\epsfbox{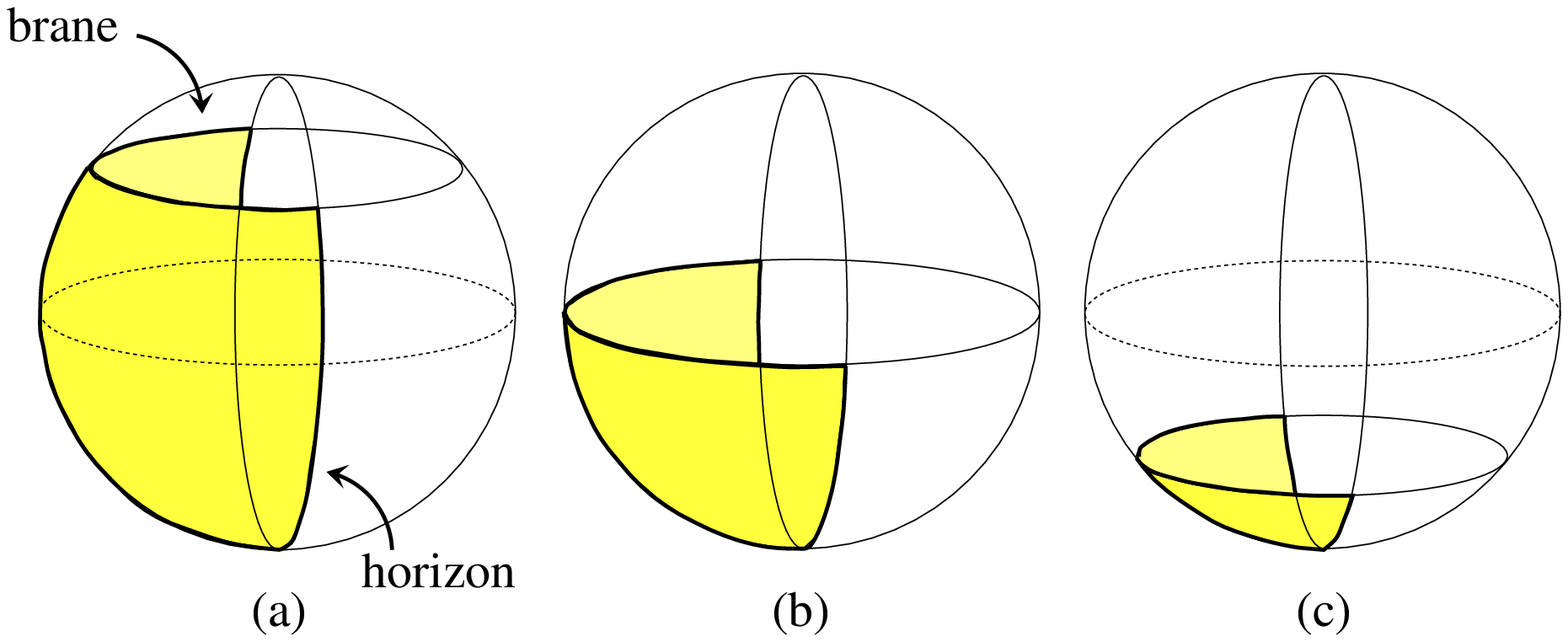}}

\ifig\spatial{A schematic picture of the spatial geometry of the
static patch at $t={\rm const.}$ On each side of the brane the
spatial curvature of the horizon is different. The branes are not
really straight, as can be seen from \hemispheres.  Note
that as the tension of the branes decreases (moving
from right to left in the sequence depicted here), they approach
a part of the horizon and the static patch approaches that of de
Sitter space ($dS_R$).}
{\epsfxsize3.8in\epsfbox{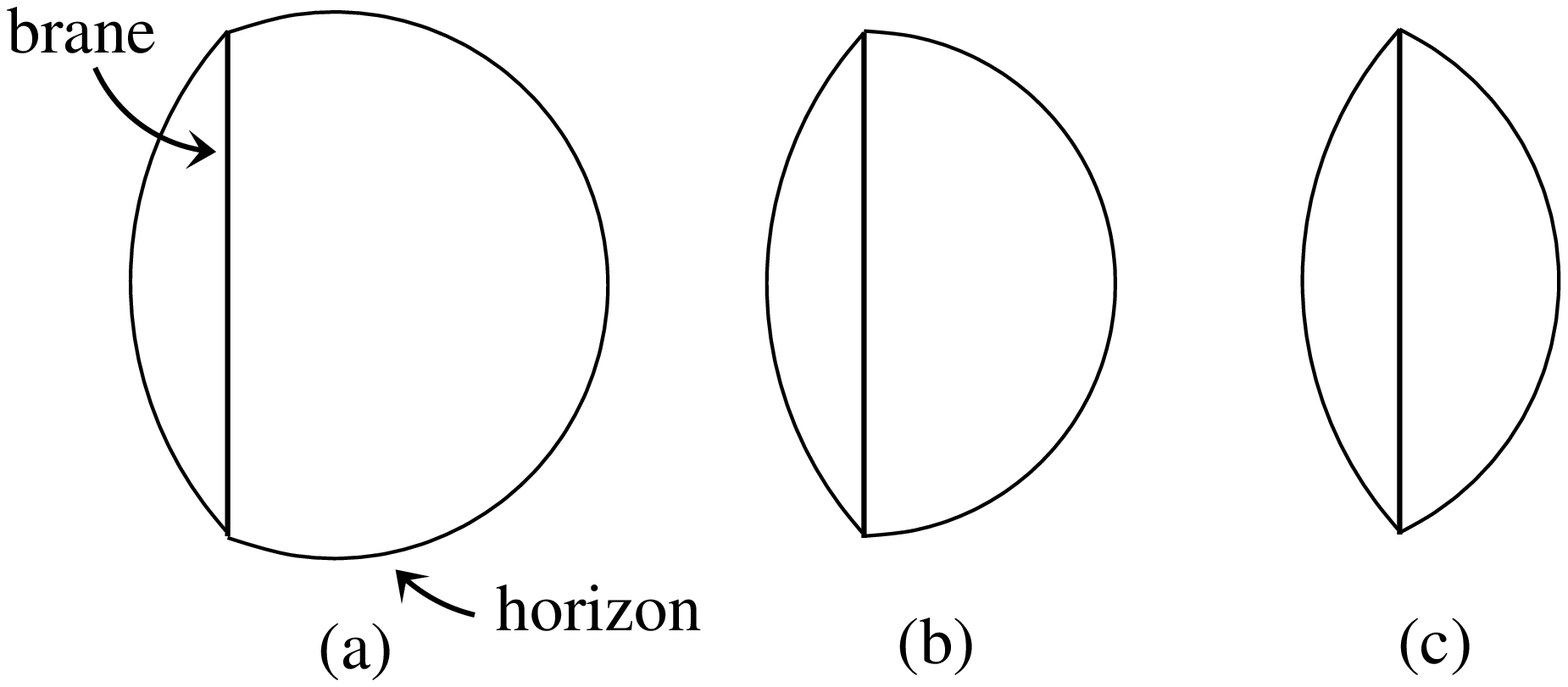}}

Notice that viewed from the embedding space, the brane is
`perpendicular' to the horizon: the brane is at $\tilde y_{d-1}  =
{\rm const.}$, whereas the horizon is at $\tilde y_d  =0$. In
other words, the brane is at a line (or surface) of a constant
latitude, whereas the horizon has a fixed longitude, namely
$\phi_{d-1} = 0^{\rm o}$ or $\phi_{d-1} = 180^{\rm o}$. The
resulting situation is depicted in \hemispheres, and a
schematic picture of this spatial slice is depicted in \spatial.

We have determined how the static patch looks on one side of the
domain wall. Of course, on the other side the situation will be
qualitatively the same, so the only task remaining is to find how
these static patches fit into the global geometry of the D-Sitter
space.

\ifig\embedding{Embedding of the static patch into the global
geometry of D-Sitter space. Each `cylinder' represents a  region
of a definite cosmological constant in D-Sitter space, and its
boundary is the location of the brane. The static patch is the
causal patch of the observer whose worldline is indicated by the
dotted line. The full global geometry of the D-Sitter (with a full
static patch embedded in it) corresponds to two diagrams of this
type, one for each side of the brane.}
{\epsfxsize5.3in\epsfbox{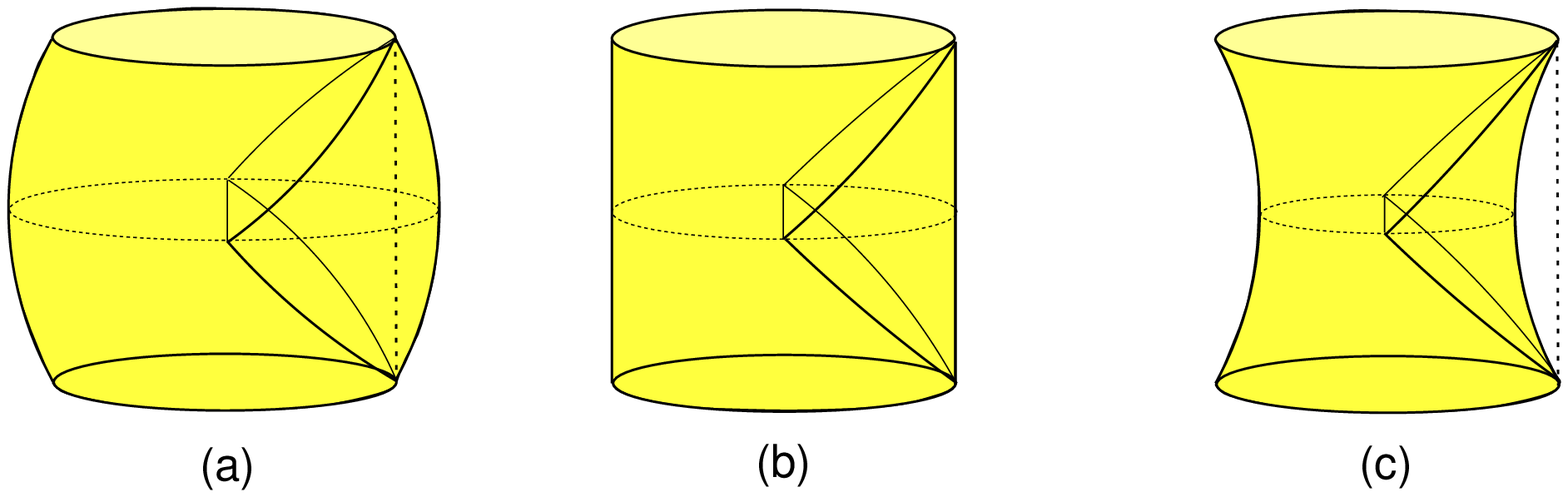}}

This question is however very easy to answer. Any observer staying
inside one particular static patch never looses  causal contact
with the brane. For this reason in the asymptotic past and the
asymptotic future the observer's worldline must come to the same
point in the Penrose diagram (see \embedding) as some part of the
brane itself. (This does not mean that the observer must actually
touch the brane, a sufficient condition is to stay a finite
distance from the brane at all times.) The whole static patch can
be therefore interpreted as the causal patch of some definite
observer. (Of course, the causal patch cannot be larger than the
static patch because it is impossible to return from behind the
horizon.) The embedding of the static patch into the global
D-Sitter is depicted in \embedding.

\ifig\babysitter{The Penrose diagram of de Sitter space for $d=2$.
Note that this Baby Sitter has rather special properties since
$S^0 = \IZ_2$. The two vertical edges of the diagram should be
identified because the spatial slice of the global geometry is a
circle. The picture shows  one static patch, which is the causal
patch of an observer originating in $A$ and arriving at $B$, and
another static patch, which is the causal patch of someone
starting at $A$ and going to $C$. All the static patches are
equivalent up to actions of the $SO(1,1)$ de Sitter group.
 }
{\epsfxsize3.5in\epsfbox{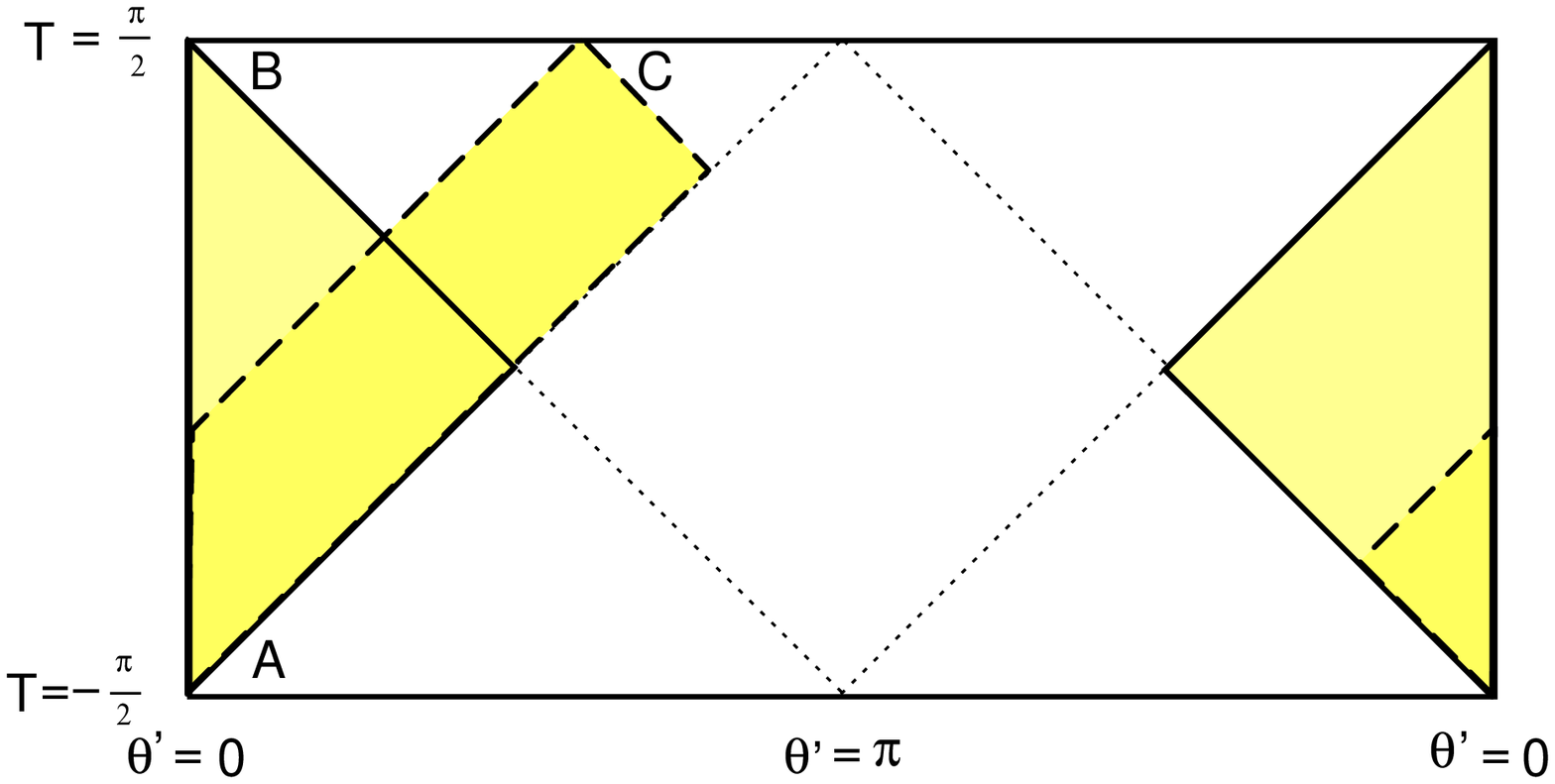}} 

For completeness, we include also  figures of the static patches
 two dimensional de Sitter and D-Sitter. We should
remember, however, that this case is rather special because $S^0$
is not a manifold, but simply just two points.

\ifig\babyembedding{The shaded square in this figure shows a
nontrivial static patch in a two-dimensional D-Sitter space. There
are also trivial static patches, the causal patches of observers
living close to $\theta'=\pi$, which do not contain the brane at
all.}
{\epsfxsize4in\epsfbox{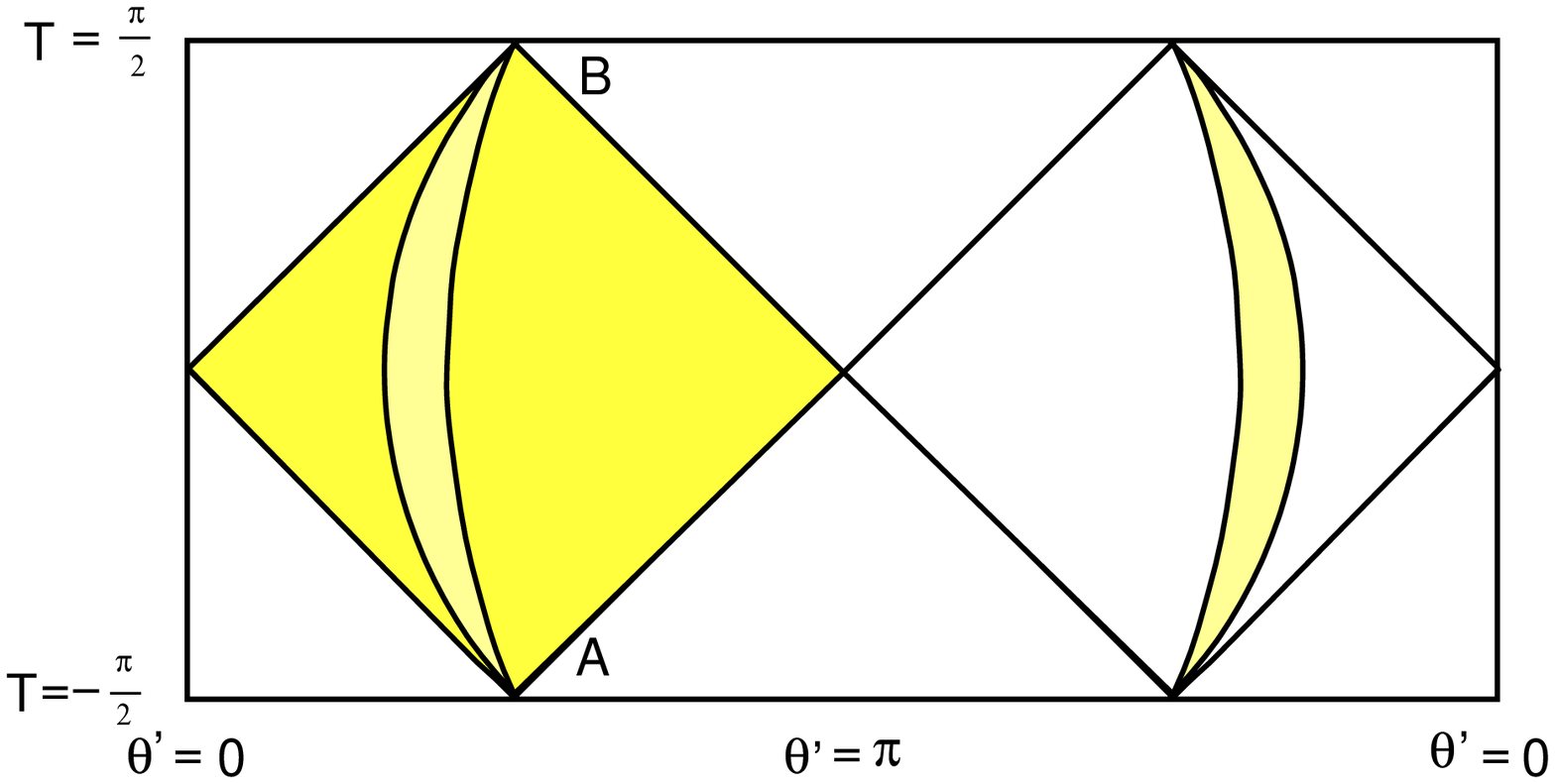}} 

\bigskip

\bigskip

In the next section, we will study the thermodynamic properties of
the brane worldvolume theory.

\newsec{Thermal equilibrium of the domain walls. }

We will be interested in the thermodynamics of our domain walls
separating  two regions of different cosmological constants. One
of the first questions that arises is whether they are in thermal
equilibrium. Naively, one might expect that this would not be the
case because the domain walls are separating  regions of different
de Sitter temperatures. We will see, however, that there is a
subtle interplay between the de Sitter expansion and the
acceleration of the branes leading to a well-defined temperature
which is constant in time.

Let us first consider the most simple case in which  the
cosmological constant in the left part of the spacetime vanishes,
and the bubble is critical (\bubbledesitter (b)),
$R_B=R_R$. We would like to know
what will be the response of a particle detector located at a
fixed position on the expanding (or shrinking) domain-wall. From
the point of view of the right part of the spacetime, the detector
is following a geodesic in de Sitter space, and it will see a
constant temperature given by the de Sitter radius $\CT = {1 /
2\pi R_R }$. From the point of view of the left part of the
spacetime (which is flat), the worldline of the detector is no
longer a geodesic. Instead, it is a hyperbola corresponding to a
constant proper acceleration $a = 1 / R_B =  1 / R_R$. For this
reason, the detector should register a constant Rindler
temperature given by $\CT ={a / 2\pi } =  {1 / 2\pi R_R }$. We see
that the two temperatures agree, and the detector can be in a
thermal equilibrium at the temperature $\CT = {1 / 2\pi R_R }$.

Similarly, the observer ${\cal O}_L$ in the middle of the brane
(\horizon)
observes a thermal bath similar to that arising from the moving
mirror problem as long as the reflection coefficient at the
transition between the two cosmological constants is nonzero.
We will analyze the constraints required to avoid back reaction
from this effect in flux compactifications in \S5.5.\foot{Thanks
to S. Hellerman and S. Shenker for discussions on this point.}

A very simple result can be obtained also for general D-Sitter
spacetimes. One way to argue for this is to consider the Euclidean
version of the geometry.\foot{We will consider only the Lorentzian
vacuum whose Green's functions can be obtained by an analytic
continuation of the Euclidean Green's functions.} The Euclidean
analog of the detector trajectory is a circle of radius $R_B$, and
for each detector
there is a $U(1)$ isometry of the solution which generates
motions along this circle. For this reason the propagator
$G(\Delta s)$ between two points on the circle separated by a
distance $\Delta s$ along the circle will have a singularity for
every $\Delta s = 2  \pi n R_B$, $n \in \IZ$. Translated into the
Lorentzian geometry this means that the propagator $G(\Delta
\tau)$ between two points on the detector trajectory (separated by
a proper time interval $\Delta \tau$) will have singularities for
imaginary $\Delta \tau$ whenever $\Delta \tau = 2 \pi i n
R_B$, $n \in \IZ$. It can be shown that the presence of these
singularities implies that the detector will register a thermal
radiation at temperature
\eqn\temperature{\CT = {1 \over 2\pi R_B}.}
(For more details, see for example the discussion related to
equations (3.58) and (3.67) of \BirrellIX.) Because in this
derivation we did not need to specify whether we think about the
detector as being on the left or on the right side of the domain
wall, it is clear that the detector can be in thermal equilibrium.
The corresponding equilibrium temperature is given by
\temperature.

This temperature \temperature\ is what the brane observer measures
locally.  In general, the effective temperature in general
relativity depends on position and on the observer making the
measurement due to blueshifting effects arising from nontrivial
warping by $g_{00}$. As we discussed in the previous section, the
brane observer in the full geometry has a static patch with
effective temperature diverging at the horizon, leading to an
entropy that depends on position on the branes which is dominated
by the region near the horizon and is as difficult to calculate as
that of the original de Sitter static patch due to cutoff
dependence.

However, other observers such as the left observer ${\cal O}_L$ (\horizon)
can probe entire spatial slices of the D-brane worldvolume by
sending out probes in all directions toward the bubble wall.
Suppose the left observer ${\cal O}_L$ sends a spherically
symmetric probe to the branes which will reach the branes around
the time $\tau=0$. Then because the branes do not coincide with
the horizon for the left observer, $g_{00}\sim 1$ there, and
${\cal O}_L$ observes the brane degrees of freedom at the local
brane temperature \temperature\ and carrying the corresponding
extensive brane worldvolume entropy to a good approximation.  In
the next section, we will calculate this entropy in our D-Sitter
spacetimes.

\newsec{D-brane Entropy as probed by  observer ${\cal O}_L$ in critical D-Sitter}

Given the well-defined thermodynamics we have developed in the
previous section, we can now study the entropy carried by the
D-branes in the D-Sitter spacetimes obtained from flux
compactifications. We have in mind the models \KachruAW\MaloneyRR\
whose relevant properties we will collect here and in \S5.2.  As
we will see, the input from the models is relatively simple and
will apply rather generally.  In this section, we will focus on
the entropy accessible to probes sent by the observer ${\cal O}_L$
in the middle of the bubble.  (In \S7, we will discuss the open
string physics of the static observer ${\cal O}_s$.)

For simplicity we will here consider critical bubbles with
flux quantum numbers (i.e. D-brane charges) $\vec Q$  separating a
phase of $\Lambda_R\sim 1/R_R^2$ with flux quanta $\vec Q_R$  from
a phase of smaller cosmological constant $\Lambda_L\sim 1/R_L^2$
with flux quanta $\vec Q_L=\vec Q_R+\vec Q$ such that
$\Lambda_R-\Lambda_L$ is of order $\Lambda_R$.  This includes the
possibility of $\Lambda_L$ tuned to approximately zero.

Here we are organizing the quantized RR fluxes into a vector $\vec
Q_\gamma$ (where $\gamma$ refers to the flux vacuum of interest,
so that e.g. $\gamma=R(L)$ refers to the right (left) de Sitter
vacuum respectively in the D-Sitter spacetimes introduced in \S2).
The kinetic term for the RR fields, $\int F\wedge *F$ then
determines the leading $\vec Q_\gamma$-dependence of the potential
energy in a given flux model
\eqn\Qprod{
\vec Q_\gamma^2\equiv Q_\gamma^2\equiv \int (F\wedge *F)_\gamma
}
We will assume that the moduli other than the dilaton are
stabilized near order one, putting in some order 1 fudge factors
to represent their effects, and we will focus on the dilaton
dependence.  So on a torus or toroidal orbifold model
\refs{\MaloneyRR,\KachruHE,\FreyHF} one would obtain simply $\vec
Q_\gamma^2\sim \sum_i q_i Q_i^2$ with $q_i$ of order 1, but on a
general Calabi-Yau the structure will be more complicated. In our
application we will be interested in the gross scaling of various
quantities with the magnitudes $Q_\gamma$.

Clearly there are many interesting ways to deviate from and refine
these choices, but we will see that these specifications are
sufficient to answer the most basic questions we are interested in
regarding the entropy comparisons. In the following, the symbol
$\sim$ will refer to relations that hold up to coefficients that
are of order one, by which we mean coefficients which do not go to
zero or infinity as we scale up the RR flux quantum numbers.

\subsec{Model Independent Analysis}

Let us begin by determining the entropy without inputting any
information about the flux stabilization of the dilaton in the
concrete models.  Then in the next subsection we will add the
constraints from the flux stabilization mechanism.  A horizon
sized bubble has a tension $T_H$ satisfying \BrownKG\
\eqn\horsized{
T_c^2={2(d-2)(\Lambda_R-\Lambda_L)\over{(d-1)l_d^{2d-4}}}
}
where $l_d$ is the $d$-dimensional Planck length. Given our
specification that $\Lambda_R-\Lambda_L\sim {\cal O}(\Lambda_R)$,
we can simplify \horsized\ to $T_c^2\sim\Lambda_R/l_d^{2d-4}$.

The tension of the $D-(d-2)$-branes which form the bubble wall
is\foot{Here the $d-2$ refers to the spatial Neumann directions of
the D-branes in the de Sitter dimensions; the branes wrap cycles
in the internal space whose dependence we are suppressing here.}
\eqn\Dtension{T_D= {\alpha Q\over g_sl_s^{d-1}}\sim {\alpha Q
g_s^{{d\over d-2}} \over l_d^{d-1}} }
where $\alpha$ is a fudge factor meant to indicate the order one
parameters that affect the D-brane tension;\foot{In
the supercritical MSS models \MaloneyRR, it happens that
$\alpha$ is significantly smaller than
one for models with a cosmological constant
tuned to be much smaller than string scale, so for
such cases we will have in mind the KKLT models
\KachruAW.}  in the second form we
have used the relation $g_s^2l_s^{d-2}\sim l_d^{d-2}$.

Plugging \Dtension\ into \horsized\ leads to the following relations.
\eqn\stringunits{
{R_R\over l_s}\sim {1\over{\alpha g_s Q}}.
}
The entropy $S_R$ associated with the
$dS_R$ horizon is given
by
\eqn\entropy{
S_R\sim {R_R^{d-2}\over l_d^{d-2}}\sim {Q^2\over{\alpha^{d-2}(g_sQ)^d}}
}
We are interested in understanding the entropy carried on the
D-branes.  The $(d-1)$-dimensional D-brane field theory has an
effective 't Hooft coupling constant which runs with energy scale;
evaluating it at the scale of the temperature ${\cal T_R}\sim
1/R_R$ we found for critical bubbles in \S4\ gives (using
also \stringunits)
\eqn\geff{ g_{eff}^2\sim {Q g_{YM}^2\over {\cal T}_R^{5-d}}={Q
g_s\over ({\cal T}_R l_s)^{5-d}}\sim {\alpha^{d-5}(Qg_s)^{d-4}. }}

In order to proceed, we need
to understand the range of couplings of interest to
us.  From \stringunits, we see that if we confine ourselves to the
region
\eqn\saneregion{ {R_R\over l_s}\ge 1 }
then we have
\eqn\saneregionII{ \alpha g_sQ\le 1 }
For $d>4$, this means (from \geff) that the effective 't Hooft
coupling $g_{eff}$ is $\le 1/\sqrt{\alpha}$, and we can apply
perturbative field theory.

For $d=4$, the effective 't Hooft coupling satisfies $g_{eff}\sim
1$, so we have to look more closely to determine whether we can
reliably study the physics of the D-brane.  The effective
Yang-Mills coupling itself (as opposed to the 't Hooft coupling)
satisfies
\eqn\gYMefffour{ {(g_{YM}^{(d=4)})^2\over {\cal T}_R} \sim {1\over
Q} << 1}
using \stringunits.

We believe this is enough to give control, as is suggested by the
work on nonconformal versions of AdS/CFT \ItzhakiDD. For $d=4$ our
D-brane is effectively a $D2$-brane (if we suppress the compact
dimensions), for which the Yang-Mills coupling runs to zero in the
UV and becomes strong in the IR. The $D2$-brane solution has three
distinct regions as one moves radially (corresponding to the
worldvolume RG flow).  Near the boundary, the solution is highly
curved corresponding to the UV free Yang-Mills theory.  At
$g_{eff}$ of order 1, it transitions to a weakly curved region
accessible to supergravity analysis. For the ordinary $D2$-brane
of type IIA theory, it is not until the effective Yang-Mills
coupling $g_{YM}^2/energy$ itself becomes very large that the
solution crosses over to the M2-brane solution. We see from
\gYMefffour\ that since we are interested in $Q\ge 1$ (mostly
$Q>>1$ in fact), $g_{YM}^2/energy \ \ <<1$ and we should expect
results similar to those in the two $D2$-brane regions.

Of course our D-branes are not literally type IIA D2-branes.
Microscopically, they are IIB D5-branes wrapped on three-cycles of
a compactification manifold in the critical geometric KKLT models
of \KachruAW, and noncritical D-branes of various dimensions
wrapped on cycles of the asymmetric orientifold in the MSS models
\MaloneyRR.   They are codimension one objects in the de Sitter
directions.  In general, we expect to have distinct $\alpha'$ and
$g_s$ expansions, which for D-brane solutions are controlled by
the effective 't Hooft coupling $g_{eff}$ and $(g_s Q)/Q$
respectively.  The large $Q$ expansion is good when the dilaton
$g_s$ is sufficiently weak even if $\alpha'$ effects are
nontrivial. This is true for $g_{eff}\sim 1$ if $Q$ is large,
which is our situation \gYMefffour.

In this regime, in the large $Q$ limit the entropy on the D-branes
is of the form
\eqn\DentropyI{ S_D=f(g_{eff})Q^2 R_B^{d-2}{\cal T}_B^{d-2} }
where ${\cal T}_B\sim {1\over R_B}$ is the D-brane temperature
\temperature.  (Again, it is worth emphasizing that here we
are discussing the extensive entropy on the D-branes accessible to
measurements performed by the observer ${\cal O}_L$, who
is in causal contact with all points on the branes and sees
the extensive D-brane entropy \DentropyI.)

Using  this relation, we can simplify \DentropyI\ to
\eqn\DentropyII{ S_D=f(g_{eff})Q^2 }

For future reference we also note here that from \ItzhakiDD\ the
proper thickness of the brane solution
is of order
\eqn\thickness{ L_{branes}\sim l_s g_{eff}^{1\over 2} }
(Of course this should not be taken literally for $g_{eff}<1$
since GR will break down in this regime; in this regime we should
expect an effective thickness of order $l_s$.)

From these relations, we see that if $f(g_{eff})$ is of order one,
the D-brane entropy $S_D$ is
\eqn\Dentropynaive{
S_D\sim Q^2,
}
and its relation to the $dS_R$ entropy $S_R$ is
\eqn\inequ{
S_D << S_R ~~~~~~{\rm for}~~~~~ R_R >> l_s
}
and
\eqn\corrpt{
S_D\sim S_R ~~~~~~~~{\rm for}~~~~~~ R_R\sim l_s
}
This latter limit is obtained if
\eqn\gcorr{
g_s\to 1/Q
}
which in particular means that $g_{eff}$ approaches order one. In
this same limit, the thickness of the brane solution \thickness\
approaches $l_s$.  $f(g_{eff})$ is of order 1 for $g_{eff}\sim 1$
in many D-brane systems studied, so we believe this is a
reasonable assumption.

We will refer to this limit \corrpt\ as the ``correspondence
point'', and will discuss in \S5.6\ its relation to the
usual correspondence point for black hole physics
\refs{\SusskindWS,\HorowitzNW}.

In \S5.3\ we will find that
in the flux stabilization models the correspondence point is
achievable and arises when $\Lambda_R$ is of order the maximum
classically stable cosmological constant available in the models.

Applying the formula \simpleA\ for the area of the horizon in the
left observer ${\cal O}_L$'s causal patch, in the case of a nearly
flat bubble $R_L>>R_R$, we can obtain a precise relation
expressing the tradeoff in entropy observed by
${\cal O}_L$ between the horizon and the
bubbles. Combining \simpleA, \Dtension, and \Dentropynaive, we
obtain
\eqn\SAsum{ {R_L^2 \over R_R^2} \left({A\over
|\Omega_{d-2}|}\right)^{{2\over d-2}}+R_B^2R_R^2{\alpha^2
g_s^4\over l_4^2 (d-2)^2}S_D=2(2R_R^2-R_B^2) }
This expresses a tradeoff between horizon entropy and brane
entropy for ${\cal O}_L$ at $\tau=0$.

\subsec{Model Input: Generalities}

In this subsection we will collect some of the relevant details
from the flux stabilization models leading to $dS$ vacua. These
models make use of contributions to the moduli potential--coming
from fluxes (at order $g_s^0$ in string frame), orientifolds and
other sources of negative tension arising at order $g_s^{-1}$ in
string frame, and positive contributions to the potential coming
at leading order $g_s^{-2}$ in string frame--to stabilize the
dilaton.  All the moduli are stabilized by this type of mechanism,
by orbifolding, or by perturbative and nonperturbative quantum
corrections to the $d$-dimensional effective potential.

Focusing on the dilaton dependence, the cosmological term to the
first three orders in the $g_s$ expansion is of the form
\MaloneyRR\
\eqn\dilpot{ \Lambda_\gamma(g_s)={g_s^{{4\over d-2}}\over l_d^2}
\biggl(a-b_\gamma g_s+{b_\gamma^2\over
4a}(1+\delta_\gamma)g_s^2+{\cal O}(g_s^3)\biggr) }
where the subscript $\gamma$ refers to which $RR$ flux and brane
configuration has been chosen.  The third term comes from the
kinetic terms $\int F_{RR}\wedge * F_{RR}$ for the RR field
strengths $F_{RR}$, so
\eqn\cterm{{b_\gamma^2\over 4a}(1+\delta_\gamma)\sim Q_\gamma^2. }
The other two terms have the following origin in the microscopic
models.  For the KKLT models \KachruAW, $a\sim Q_{NSNS}^2$ since
the $H_{NSNS}$ kinetic term arises at order $1/g_s^2$ in string
frame; $b\sim {\chi\over 24}-N_3-N_{\bar 3}$ where $N_3, N_{\bar
3}$ are the numbers of D3-branes and anti-D3-branes and where
$\chi$ is the Euler character of the Calabi-Yau fourfold in F
theory since the crucial negative term in the potential arises
from the contribution of wrapped D7-branes which contribute
negative D3-brane charge and tension of this order \GiddingsYU.
This latter contribution can be related to the flux background via
a Chern-Simons contribution to Gauss' law \GiddingsYU, giving
\eqn\gausslaw{ {1\over 2 g_s^2l_s^8T_3}\int_{M_6}H_{(3)}\wedge
F_{(3)} =-Q_3^{localized}}
where $Q_3^{localized}$ is the 3-brane charge coming from all
localized sources (D3-branes, orientifold planes, and wrapped
D7-branes), $T_3$ is the D3-brane tension, and $M_6$ is the base
of the F-theory fourfold compactification.

 For the MSS
models \MaloneyRR, $a\sim D-10$ where $D$ is the total dimension
of the supercritical theory, and $b$ comes from orientifold and
antiorientifold planes and is independent of the flux quantum
numbers.

In both types of models, we will take $a$ to be finite as we scale
up the RR charges, and thus will include it in the ``order 1"
coefficients we are not keeping track of.

The parameterization \dilpot\ is useful because as noted in
\MaloneyRR, $\delta=0$ corresponds to a flat solution
$\Lambda_\gamma=0$, and for $0<\delta<{(d-2)^2\over 8d}$ one finds
metastable de Sitter minima with cosmological constants ranging
from zero to
\eqn\lammax{{\Lambda_{max}\sim {1\over l_d^2}{{a^{d+2\over
d-2}8^{4\over d-2}(d-2)^2}\over {b^{4\over d-2}d(d+2)^{d+2\over
d-2}}}}.}
In this range, the dilaton does not vary much, changing from ${2a/
b}$ to ${8a/ (b(d+2))}$.  From \cterm\ and the just quoted range
of $\delta_\gamma$, we see that if we just keep track of the
$Q_\gamma$ dependence
\eqn\bQ{ b\sim Q_\gamma }
in any $\Lambda\ge 0$ minimum.  In particular, since
$0<\delta<{\cal O}(1)$ for the whole range of models, $Q_\gamma$
is of the same order for all $\gamma$ in the discrete family of models for
which $b$ is independent of $\gamma$. This means that the string
coupling
\eqn\basicgfluxII{ g_s\sim {1\over Q_\gamma} }
is of the same order in all of the models (in particular it does
not change much as we go from one side of the D-brane domain walls
to the other side). Using these relations, we see that
\eqn\lammaxstring{ \Lambda_{max}\sim {1\over l_s^2} }

Using \cterm\ for both $dS_L$ and $dS_R$, we obtain a condition on
the D-brane charges required for a domain wall separating $dS_L$
and $dS_R$ if $b$ does not change in the transition:
\eqn\Qcondition{
\vec Q^2-2\vec Q\cdot\vec Q_L = {b^2\over 4a}(\delta_R-\delta_L)
}

In general, this condition has many solutions depending on how
$\vec Q$ is oriented relative to $\vec Q_L$.  This set of
solutions in general will include a range of $Q\equiv |\vec Q|$.
This range is bounded above by $Q_R\sim Q_L$, since for $Q$ much
larger than this, the condition \Qcondition could not be
satisfied. So for the D-brane walls in flux models, we have
\eqn\Qbound{ Q\le Q_L\sim Q_R }
(Here we are using the fact that \cterm\ implies that $b$ is of
the order of $Q_L\sim Q_R$ in the range of $\delta$'s we have
discussed above for which dS minima exist.)

In the KKLT models, there are important other ingredients going
beyond the terms so far listed in \dilpot, which stabilize the
volume modulus $\rho$ of $M_6$ and which also shift the minimum of
the potential in the dilaton direction.

KKLT start from a no scale model \GiddingsYU\ with complex
structure moduli and dilaton stabilized by a potential which in
components takes the form \dilpot\ in each direction in the scalar
field space. This no-scale potential $\Lambda_{ns}$ satisfies
$\Lambda_{ns}\ge 0$, with $\Lambda_{ns}=0$ solutions given by
setting $D_iW_0=0$ where $i$ runs over moduli other than $\rho$
and where $W_0$ is the classical superpotential given by \GukovYA\
\eqn\superpot{ W\sim \int_{M_6}\Omega\wedge (F_3-\tau H_3) }
Here $\tau=C_{(0)}+ig_s^{-1}$ is the axion-dilaton and the
superpotential is being evaluated at an orientifold limit of the F
theory compactification where $M_6$ becomes an orientifold of a
Calabi-Yau with holomorphic three-form $\Omega$.

From our component expression \dilpot\ and the ensuing discussion,
we see that
\eqn\bnoscale{ b=2Q_{NS}Q_\gamma ~~~~~~~{\rm
for}~~~\Lambda_{ns}=0. }

If we start from such a $\Lambda_{ns}=0$ vacuum and change the RR
flux $F$ (as we with to do using D-branes in our present
application), there are two basic cases to consider.  One case,
appropriate to the discussion surrounding \Qcondition, is to
consider a jump in flux for which $b$ stays constant and to take a
model we obtain on the other side of the domain wall for which at
the no-scale level $\Lambda_{ns}>0$.  Another possibility is to
stay within the set of models given by $\Lambda_{ns}=0$, which
requires $a$ and $b$ to change appropriately to preserve
\bnoscale. For flux jumps in which $\int H\wedge F$ changes, there
must be D3-branes or anti-D3-branes ending on the D5-brane domain
wall in order to preserve the relation \gausslaw.

In fixing the volume modulus $\rho$, KKLT employ a no-scale
violating contribution to the superpotential such as a gaugino
condensate to obtain an AdS minimum, and anti-D3-branes to kick
the minimum up to a dS minimum.  These additions roughly add
\eqn\newterms{ \Delta \Lambda\sim -e^K|W_0|^2+T_{\bar 3} }
to the scalar potential, where $T_{\bar 3}$ is the tension of the
anti-D3-branes in Einstein frame, and where in the first term we
have used the fact that the solution to the equation of motion for
$\rho$ balances the gaugino condensate superpotential against
$W_0$, leading to a negative contribution of the order of the
first term in
\newterms \KachruAW.  It may be possible to eliminate the need for the
anti-D3-brane contribution to $\Lambda$ by making use of
$\Lambda_{ns}>0$ metastable flux vacua in the no-scale model
instead of starting from $\Lambda_{ns}=0$ configurations and
adding anti-D3-branes. We leave a detailed investigation of that
for future work.

There is a rich set of flux jumps available in this set of
models.\foot{See \freyetal\ for an analysis of decays arising from
brane bubbles in these models.}  We will mostly confine ourselves
here to verifying that the correspondence limit \corrpt\ we found
in the model independent analysis is available in flux stabilized
models.

\subsec{Model Input:  Correspondence Point}

We would like to understand whether the correspondence point
\corrpt\ at which the D-brane entropy approaches the $dS_R$
entropy is available in the concrete flux models. This requires
simply that
\eqn\corrI{ \Lambda_R-\Lambda_L\sim \Lambda_R\sim{1\over l_s^2} }
and that there exists a $Q\sim Q_L\sim Q_R$ critical bubble.

Consider now the case where $b$ does not change in the flux jumps,
and where $\delta_L$ is approximately zero,\foot{We will explain
how fine the discretuum is more precisely in a future subsection.}
and $\delta_R$ is of order 1. This means $\Lambda_R$ scales like
$\Lambda_{max}\sim {1\over l_s^2}$.  Then to have a critical
bubble, we need $T_D^2\sim l_s^{-2}l_d^{2d-4}$ (from \horsized),
which implies $g_s\sim 1/Q$.

This is satisfied here, as we can see as follows.
The condition \Qcondition\ becomes
\eqn\Qcondmax{ Q^2-2\vec Q\cdot\vec Q_L\sim {b^2\over 4a}\sim
Q_L^2 }
which means that $Q$ must be of order $Q_L\sim Q_R$.
So in this case we have
\eqn\maxcase{
b\sim Q\sim Q_L\sim Q_R.
}
Then since at the dS minima $g_s\sim 1/b$, and since $b \sim Q_L$,
we have from \maxcase\ that $g_s\sim 1/Q$.

So in this case, \gcorr\ is satisfied, and we are at the string
scale correspondence point \corrpt\ where the D-brane entropy
$S_D\sim Q^2$ is of the same order as the $dS_R$ entropy $S_R\sim
Q_R^2$.

In the KKLT models, we can also obtain the correspondence limit in
the cases where $b$ does change in the flux jumps.  For example,
we can consider jumps which preserve $\Lambda_{ns}=0$ at the
no-scale level, and change $W_0$ so as to adjust the negative term
in \newterms.  Again, it is possible to start with a nearly string
scale cosmological constant and jump the flux by a $Q\sim Q_R$
D-brane bubble, which means as above that $g\sim {1/Q}$ and the
bubble is critical given our string-scale starting point.

We have considered critical bubbles in formulating and
exhibiting our correspondence point.  One might wonder about the
possibility of subcritical bubbles also saturating the $dS_R$
entropy in some situation.  In fact, we can rule out this
possibility for $R_R\ge l_s$ from the basic feature $g_s\sim
1/Q_\gamma$ in the models.  Suppose the possibility existed, which
means we could start from $\Lambda_R\le 1/l_s^2$ with a subcritical
bubble saturating the $dS_R$ entropy.   Then increase $Q$ to
obtain a critical bubble (fixing $\Lambda_R$).  Since this
increases the D-branes' entropy $S_D$, it would lead to a
contradiction with the result \inequ\corrpt\ which show that for
critical bubbles the entropy is subdominant unless
$\Lambda_R$ is of order the string scale.  We must check that this
process of increasing $Q$ to obtain a critical bubble is
possible (i.e. that $Q$ is not somehow constrained to provide only
subcritical bubbles in any case). In the models, we can consider
$Q$ up to of order $Q_\gamma$ as we discussed in \Qbound.
 The upper limit $Q\sim Q_\gamma$ is not consistent with a
 subcritical bubble, since such a bubble would satisfy
 $T_D^2<(\Lambda_R/l_d^2(d-2)$ which would imply
 $R_R/l_s<1/(Q_\gamma g_s)$.  But since $g_s\sim 1/Q_\gamma$, we
 see that this would require $R_R<l_s$ which is outside the regime
 of validity of the analysis.  As we will see in \S6, this
 simple relation $g_s\sim 1/Q_\gamma$ coming from the flux models
 also guarantees satisfaction of Bousso's bound.

\subsec{Model Input: more generic D-Sitter spacetimes}

If $\delta_R << 1$, we find that there exist cases with
critical bubbles for which $S_D << S_R$, as well as a rich spectrum
of sub and supercritical bubbles.

Let us begin by estimating how finely spaced the discretuum is in
our situation with many fluxes, following Bousso and Polchinski
\BoussoXA. If the space of fluxes is $\chi$-dimensional, then
\eqn\cdef{\rho^2\equiv {b^2\over 4a}(1+\delta_\gamma)\sim
\sum_{i=1}^\chi q_iQ_{\gamma,i}^2 }
where $q_i$ are coefficients we are assuming to be of order 1.
This defines a $\chi-1$-sphere of radius $\rho$.  Let us consider
a shell between $\rho=Q_\gamma$ and $\rho=Q_\gamma+\eta$.  The
volume of this shell is
\eqn\volshell{ \eta{{\rho^{\chi-1}2\pi^{{\chi-1\over 2}}}
\over\Gamma({\chi-1\over 2})} }
which is roughly the number of points in the shell. Setting this
to one, we obtain the discretuum spacing
\eqn\delmin{ \delta_{min}\sim {4a\over
b^2}\biggl({2\Gamma({\chi-1\over 2})\over
Q_\gamma^{\chi-2}2\pi^{{\chi-1\over 2}}}\biggr) }
This translates into a minimum positive cosmological constant
\eqn\minlam{ \Lambda_{min}\sim {1\over l_d^2}a({2a\over
b})^{4\over d-2}\delta_{min} }
with a corresponding entropy
\eqn\entlammin{ S_{\Lambda_{min}}\sim {b^{d}\pi^{{\chi\over
4}(d-2)}Q_\gamma^{{\chi\over 2}(d-2)}\over \chi^{{\chi\over
4}(d-2)}} }
Now using the fact discussed in \S5.2\ that for the KKLT models
$b\sim Q_\gamma\sim \chi$, we obtain an estimate for the entropy
in the class of examples we are considering of the form
\eqn\entsimple{ S_{\Lambda_{min}}\sim Q_\gamma^{{\chi\over 2}+4} }

If we consider the right dS , $dS_R$, to have cosmological
constant of this scale $\Lambda_R\sim\Lambda_{min}$, then the
naive D-brane entropy \Dentropynaive\ is very much smaller than
the dS entropy because of the bound \Qbound\ on the flux charge we
can pull out in the form of branes.

Let us make three
remarks about interpreting
this result. Firstly (as will also be discussed in
\S5.6) in the canonical ensemble taking into account
Boltzmann suppression of heavy states,
in this regime far from the correspondence point,
the  DS entropy as seen by ${\cal O}_L$ is less than dS entropy, even
if we include states not localized on the D-branes.
Secondly, as will be discussed in \S7, the entropy
associated with a putative open string theory at the horizon
of the static patch in
de Sitter space agrees with the entropy ascribed to the
horizon of de Sitter space even in cases where it
is very large as a function of flux quantum numbers as
in \entsimple.  Thirdly, the formula \entsimple\
appearing here and in AdS flux models may be suggestive of a
field theoretic system
with many flavors.

Using the general relations in \S5.2, we can study many D-Sitter
spacetimes.  Let us take $\Lambda_L=\Lambda_{min}$ and consider a
pair of representative set of cases for $\Lambda_R<\Lambda_{max}$.
(The case $\Lambda_R\sim \Lambda_{max}$ was studied in the
previous subsection where we saw it gave rise to a correspondence
point where the D-brane entropy is of the order of the $dS_R$
entropy.)

\noindent{\it Case I:  $(\delta_{min}, \delta_{min})$}.

First let us consider the case where $\Lambda_R\sim\Lambda_{min}$,
that is $\delta_R > \delta_L$ but both of order $\delta_{min}$.
Since $\vec Q_L^2$ and $\vec Q_R^2$ are separated by the minimal
spacing $\delta_{min}$, there is as we just discussed on average
one lattice point in the shell between $Q_R$ and $Q_L$.  This
means the average separation between $\vec Q_R$ and $\vec Q_L$,
namely $\vec Q=\vec Q_L-\vec Q_R$, has magnitude of order $Q_R\sim
Q_L$.  So for case I, $Q\sim Q_L\sim Q_R$.

We can compute the size of the corresponding bubble. For this
case, we find that the D-brane tension $T_D$ satisfies $T_D^2\sim
{Q_L^{-{4\over d-2}}}/l_d^{2d-2}>> (\Lambda_R\sim \Lambda_{min})$.
So the bubble is highly supercritical, and $R_B\sim 0$.

\noindent{\it Case II:  $(\delta_{min}, \delta_R\sim 1/Q_L)$}

Now let us consider an intermediate case in which
$\Lambda_{min}<<\Lambda_R<<\Lambda_{max}$.  If we take
$\delta_R\sim 1/Q_L$, then we find \Qcondition\ can be satisfied
for a range of $Q$ from order 1 to order $Q_L$.  For d=4, the
former is very subcritical and the latter is supercritical;
in between at $Q$ of order $Q_L^{1/2}$ we find a horizon
sized bubble.  Again, the D-brane entropy is generically very
subdominant to the $dS_R$ entropy.

\subsec{Constraint on the D-Sitter models from back reaction}

As mentioned in \S4, we must consider the back reaction of the
thermal radiation coming from the accelerating brane.  If we model
this as a ``moving mirror'' phenomenon\foot{We thank S. Hellerman
and S. Shenker for discussions on this interpretation.}  then we
obtain a temperature felt by ${\cal O}_L$ which is of order
$\rho/R_B$ where $\rho$ is the reflection coefficient for modes
impinging on the brane from the ``left" side; $\rho$ is $\le 1$ and
goes to zero as $\Lambda_L\to\Lambda_R$. The back reaction this
produces on the curvature is of order
\eqn\curveback{ \Delta {\cal R}\sim G_N \left({\rho\over
R_B}\right)^d\sim \left({l_s\rho \over R_B}\right)^d{g_s^2\over
l_s^2} }
In order to avoid back reaction, we must consider $\Lambda_L\ge
\Delta {\cal R}$.  This is easily achieved in the models.  Note
that contributions to the stress energy which simply renormalize
the cosmological term can be tuned effectively with our fluxes;
the values for $\Lambda_\gamma$ we quote in the analysis refer to
the resulting total cosmological constant.

\subsec{On the relation between D-Sitter and de Sitter entropy}

The concrete results in this section pertain most directly to
D-Sitter space.  In this subsection we will explore their relation
to de Sitter space itself.  As we have discussed, D-Sitter space
is a deformation of de Sitter space; we
have compared the D-brane entropy $S_D$ probed by ${\cal O}_L$
at $\tau=0$
with the entropy of the corresponding $dS_R$ space, and found it
to agree up to order one coefficients in the correspondence limit
\corrpt. Away from the correspondence limit, the entropy localized
on the D-brane bubble wall itself is parameterically smaller than
that of $dS_R$ \inequ.  In these cases, there is room in the
D-Sitter geometry for other states (such as black holes in the
middle of the bubble) but the most entropic of these
states are Boltzmann suppressed in the canonical ensemble
that applies at the temperature of the system.


It would be interesting to understand what the precise
relation between dS and DS entropy is.  In the AdS/CFT
cases we discussed in \S1\ and \S2, it is clear
that the deformation between AdS and ADS could be
made adiabatically (in the sense that one
could follow the states of the system even as
they mass up along the Coulomb branch);  but there also if we worked
in a canonical ensemble at fixed nonzero temperature the
entropy would decrease as we pull branes far enough out of
the horizon, since the stretched strings then become
Boltzmann suppressed.  In our closed cosmology it is
not clear a priori whether one can deform the system
adiabatically, but our result \corrpt\ on the string
scale correspondence point exhibits a
case where the deformation is adiabatic up to order one
coefficients in the canonical ensemble.


The assertion that the de Sitter horizon area itself should
correspond to an entropy \GibbonsMU\ is based on arguments such as
the following.  One is the possibility of trading ordinary matter
entropy for horizon area locally (see \BoussoJU\ for a review).
Another argument based on the static causal patch in dS notes that
the blueshifting of modes toward the horizon of the causal patch
leads to an area's worth of high temperature modes there
\DysonNT\SusskindSM\RandallTG; another way to view this is via the
large acceleration required of an observer to probe the horizon
but stay within the causal patch. (These arguments can be applied
also to the time dependent causal patches such as that we found
for ${\cal O}_L$ in \S3; one obtains different results for the
entropy associated to the horizon for experiments conducted at
different times.)

\bigskip
 \noindent{\it Observer ${\cal O}_R$ and the physics behind the horizon}
\bigskip

One intriguing aspect of the D-Sitter geometry is the following.
For the cases of subcritical or critical bubbles, the causal
patch for the right observer ${\cal O}_R$ is the same in DS and in
dS.

\ifig\xspace{Given the geometry of the static patch of the
observer on the right (at $\theta' = \pi$) to be de Sitter space
${{\rm dS}_{\rm R}}$, there are still more possibilities for the
global structure of the spacetime. For example, the left part of
the spacetime may be (a) de Sitter ${{\rm dS'}_{\rm L}}$ with the
same cosmological constant, or (b) it may contain a domain wall
and a region of a different cosmological constant.  }
{\epsfxsize5in\epsfbox{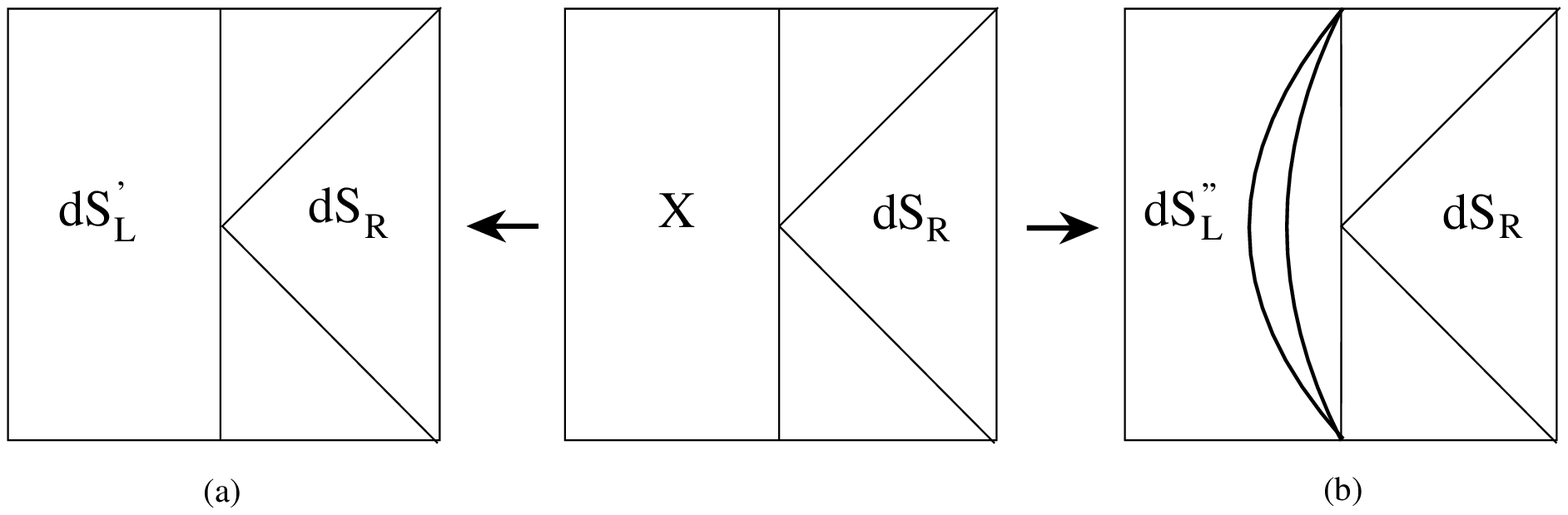}} 

That is, as illustrated in \xspace, the right causal patch
can be coupled consistently to the left side of D-Sitter or the
left side of ordinary $dS_R$.  At the perturbative level, these
are distinct possibilities for the global structure of the
spacetime.  In other words, what is behind the horizon of ${\cal O}_R$
at $\tau=0$,
which in ordinary de Sitter space is the $\tau=0$ slice
of another causal patch, can
be replaced by the left side of D-Sitter space.  Naively one
might conclude from this that the entropy of ${\cal O}_R$'s
horizon should agree with that of the left D-Sitter space,
inasmuch as this horizon entropy encodes what is behind
the horizon in a given perturbative spacetime background.
However, as we have discussed, the entropy of the left part
of DS as measured by ${\cal O}_L$ in the canonical
ensemble is smaller than $S_R$
away from the correspondence point.

In any case, up to order one coefficients our result \corrpt\ provides
prima facie evidence for an adiabatic transition from string
scale dS to DS.

\newsec{Bousso's Entropy Bound}

Bousso has proposed a bound on entropy going through light sheets
emanating inward from any surface of area $A$ in a spacetime
\BoussoXY\ (for a beautiful review see \BoussoJU).  The covariant
entropy bound is expressed in terms of light sheets, which are
lightlike hypersurfaces, emanating from a chosen surface $B$ in a
spacetime, which contract (or at least do not expand).  The
conjecture, which has been well tested in a wide variety of
circumstances and proved under some assumptions \FlanaganJP,
states that the entropy on any light sheet of a
surface $B$ in a spacetime will not exceed one quarter of the area
of $B$. As described in \BoussoJU, the conjecture is motivated by
its elegant covariance and by the desire for wide applicability of
the holographic principle. Recently a refinement to take into
account quantum effects has been proposed \StromingerBR.  One can
also formulate a related conjecture for partial light sheets
\FlanaganJP, for which the entropy going through is bounded by the
difference of the areas on the two ends. In this subsection we
will analyze both of these conditions in D-Sitter space for
two illustrative light sheets.

\bigskip
\ifig\bousso{Two light sheets which illustrate the covariant
entropy bounds
in D-Sitter space.} {\epsfxsize4in\epsfbox{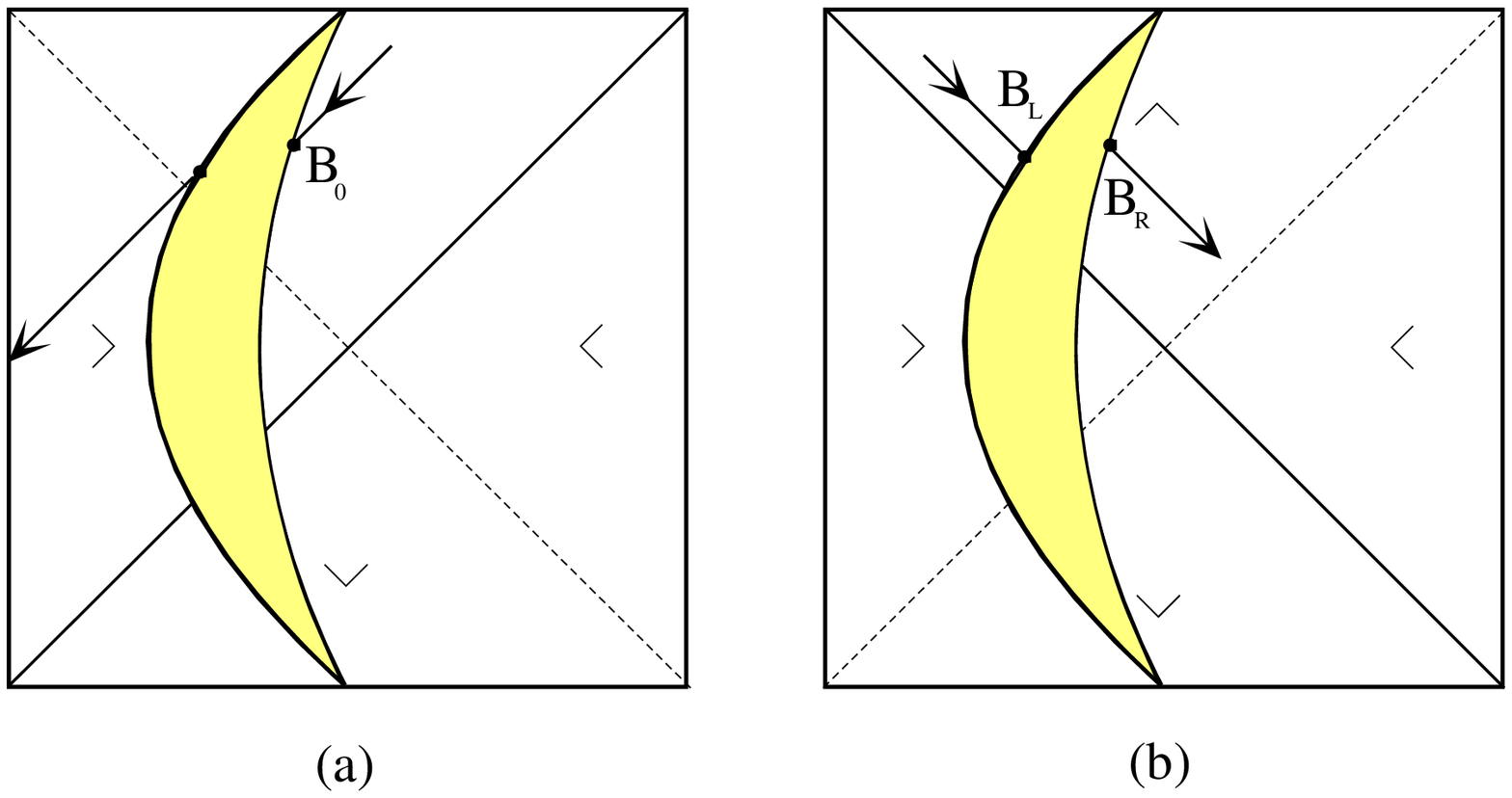}}

Let us examine first the statement for full light sheets.  In
particular, let us study what the bound predicts regarding the
entropy carried by the bubbles in D-Sitter.  If we pick a
subcritical bubble in $dS_R$, then one can consider a spherical surface
$B_0$ of area $A_0$ just to the right of the bubbles but still on
the upper left quadrant of the Penrose diagram  (see \bousso (a)); for such
surfaces $B_0$ there exists a light sheet emanating toward the
lower left in the diagram.  Let us consider this light sheet
for an example of a bound following from a full light sheet.

According to Bousso's conjecture, the entropy in this light sheet
must not exceed ${A_0/4 l_d^{d-2}}$.  This implies that
\eqn\Boussobubble{ S_D\sim Q^2 A_0{\cal T}_B^{d-2} < {A_0 \over 4
l_d^{d-2}}}

The area drops out of this relation, which can be rewritten using
$l_d^{d-2}\sim g_s^2l_s^{d-2}$ and ${\cal T}_B\sim 1/R_B$ as
\eqn\newform{ g_s^2 < {1\over Q^2}\biggl({R_B\over
l_s}\biggr)^{d-2} }
When $R_B$ is of order $l_s$, this becomes
\eqn\gcond{g_s^2 < {1\over Q^2}}
which is satisfied for flux compactifications in string theory
since \basicgflux\basicgfluxII\ $g_s\sim 1/Q_\gamma$ and
$Q\le Q_\gamma$. If
further we are at the correspondence point $R_R\sim l_s$ for a
critical bubble \corrpt, then $Q\sim Q_\gamma$ so that
$g_s\sim 1/Q$ and Bousso's
bound is saturated for all $\tau>0$.

Now let us illustrate the bound on partial light sheets going through
the bubble wall \FlanaganJP\ (see \bousso (b)). For example, we can work in the
region where both the left and right sides of the branes are in
the upper right half of the corresponding $dS_L$ and $dS_R$
Penrose diagrams. In this region, there is a light sheet with
areas decreasing as one moves down and to the right on the Penrose
diagram. Let us consider a slightly subcritical bubble. Let us
consider a surface $B_L$ to the left of the bubble with area
$A_L$, and take a partial light sheet heading down and to the
right in the Penrose diagram ending at a surface $B_R$ of area
$A_R$ just to the right of the D-branes, with all of the above
contained in the upper left quadrant. According to the covariant
entropy bound applied to this partial light sheet, we should find
\eqn\partialBousso{ {{A_L-A_R}\over 4 l_d^{d-2}} ~~>~~ S_D }
If the branes were infinitely thin, then the left hand side of
\partialBousso\ would be zero.  In our case, they have a thickness \thickness\
determined by their supergravity solution.  Let us introduce this
order of thickness in the right region of the solution (this
provides the most dangerously small contribution to the change in
area). The proper thickness \thickness\ is also the change in
$\tau$ as we go from $B_L$ to $B_R$.  Using this and the relation
\entropy\ applicable for our almost critical bubble, we
obtain
\eqn\areachange{ e^{-(d-2)\tau_R}{{A_L-A_R}\over l_d^{d-2}}\sim
{L_{branes}\over l_d}\left({R_R\over l_d}\right)^{d-3}\sim
{{(g_sQ)^{{d-4\over 4}}l_s}\over l_d}\left({R_R\over
l_d}\right)^{d-3} \sim Q^2(g_sQ)^{-{3d\over 4}}}
Then the condition \partialBousso\ becomes
\eqn\partialfin{ Q^2(g_sQ)^{-{3d\over 4}}>Q^2 }
which is satisfied in our regime $g_sQ\le 1$ (which follows from
$R_R\ge l_s$ \saneregionII).  As before, this bound is saturated
at the correspondence point.  Note that in the cases $d>4$ in
which the nominal thickness \thickness\ is substring scale, the
condition \partialfin\ is stronger than we actually need to
satisfy the bound since we should in that case take an effective
thickness of $l_s$.

\newsec{ Static observers, the DS $\leftrightarrow$ dS
deformation, and open strings}

The left causal patch of ${\cal O}_L$ (\horizon) in D-Sitter and the static
causal patch of ${\cal O}_s$ are each deformations of the causal
patch of ordinary $dS_R$. As we discussed in \S2, this deformation
is reminiscent of the deformation in AdS/CFT taking the field
theory out along its Coulomb branch, and the causal patches of the
observers ${\cal O}_{L, s}$ are reminiscent of the Poincare patch in
AdS/CFT. This is particularly appropriate as an analogy for ${\cal
O}_s$ for which there exists a static coordinate system covering
its causal patch.

In any situation with a horizon, if there is a perturbative string
description applying to a single causal patch, then since the
worldsheets have boundary on the horizon there may be a dual
channel open string description \SusskindSM\ (see also
\refs{\SenIN, \GutperleAI} for other indications of an open string
description of a time dependent background).  A priori it is not
clear how to quantize strings in such backgrounds; the nonlinear
sigma model is nontrivial and the status of the S matrix is
unclear.  Indeed, it would be difficult to formulate the problem
directly in terms of closed strings because they are in general
off shell when entering or leaving the horizon.


This situation is familiar however from perturbative string theory
in the presence of D-branes, which also inject off shell closed
strings at some locus in spacetime.  The relation to D-branes we
have developed in this paper provides support for the possibility
that a consistent open string theory may exist with the strings
ending on the horizon. One piece of evidence was discussed in \S3\
(\spatial):
the observer ${\cal O}_s$ at a fixed distance from the brane can
take a limit where the branes approach a trajectory tracking a
patch of the horizon for all time;  this is a limit in which the
static patch for ${\cal O}_s$ in D-Sitter space approaches the
original $dS_R$ causal patch.

Another piece of evidence arises from the success of the following
simple estimate for the entropy using the relation
\basicgflux\basicgfluxII\ in the flux models. Consider a de Sitter
space with horizon area $A$ coming from a string theory
compactification with flux quantum numbers $\vec Q_\gamma$. Assume
that there is a description of the system in terms of a low energy
open string theory (a field theory) with of order $Q_\gamma$
Chan-Paton factors at the horizon, corresponding to the flux
quantum numbers $\vec Q_\gamma$. Assume further that this field
theory is at a temperature ${\cal T}_{open}$ of order the string
scale (which is suggested by a picture in which the open strings
end at a stretched horizon \SusskindIF\ corresponding to a string scale
cutoff), but somewhat lower than the Hagedorn temperature
such that the divergences associated with that
transition do not come into play.  Then the entropy is given by
\eqn\Sopen{ S_{open}\sim Q_\gamma^2 A {\cal T}^{d-2}_{open}=
Q_\gamma^2 A{1\over l_s^{d-2}}={A\over l_d^{d-2}}g_s^2Q_\gamma^2 }
Now using the basic relation \basicgflux\basicgfluxII\ $g_s\sim
1/Q_\gamma$ from the flux stabilization of the dilaton, we obtain
\eqn\Sopenfin{ S_{open}\sim {A\over l_d^2}\sim S_{dS}  }
agreeing with the de Sitter entropy up to order one coefficients.

We can combine the two pieces of evidence we have gathered, and
ask for a D-Sitter space with a bubble wall of $Q_\gamma$ D-branes
and with $R_B$ of order string scale (realizing materially a
string scale stretched horizon containing D-branes).  This leads
to a $dS_R$ of order string scale, putting us again at the
correspondence point.  However, the result \Sopenfin\ on its own
seems to apply more generally.

As we mentioned in the introduction,
it is important to understand how the open string
theory we contemplate here (as well as the D-Sitter
analysis more generally) could be extended to
account for the nonperturbative decays
of de Sitter models \KachruAW\MaloneyRR\freyetal.

Finally let us note that an open string description may also apply
to non-static observers such as ${\cal O}_L$. We have exhibited a
deformation in which branes can be introduced into the left
observer's causal patch; this deformation has the property that
the horizon area at $T=0$ decreases in the process \simpleA,
suggesting again that the branes can be thought of as being pulled
out of the horizon, though in this case there is no static
description and the branes do not track the horizon in the limit
in which we go back to ordinary $dS_R$.

In summary, our results provide further evidence for the notion
that one can describe the physics of the horizon by open strings.
It is tempting to conjecture that such a description formulates
quantum gravity in
the causal patch for ${\cal O}_s$ (and maybe also ${\cal O}_L$);
however in the cases where de Sitter decays via bubble
nucleation the nonperturbative physics will not be controlled
by the simple causal patch of the perturbative de Sitter
solution \desitter.



\newsec{Conclusions}

In this paper we have studied a basic deformation of de Sitter
flux compactifications -- by introduction of D-brane domain
walls -- as an approach to the problem of exhibiting the microstates
associated to flux compactifications.  We have analyzed many basic
properties of the resulting D-Sitter spaces. In particular, we
determined the causal and thermodynamic structure of the space
from the point of view of four basic classes of observers (the
observer ${\cal O}_L$ in the middle of the bubble, the observer on
the bubble wall, the observer ${\cal O}_s$ at a fixed distance
from the wall, and the observer ${\cal O}_R$ outside a
subcritical bubble
whose causal patch is the same as that of the original de Sitter
space).

We compared the entropy localized on the D-branes in D-Sitter
space and measured by ${\cal O}_L$ to that associated with the
horizon of the original de Sitter space, and found a
correspondence point at which the entropies agree up to order one
coefficients and at which the Bousso bound is similarly saturated
for all time.

We found two pieces of circumstantial
evidence going in the direction of an open
string description of the de Sitter causal patch:  the static
observer in D-Sitter space has the D-branes approach the horizon
as one takes the limit to the original de Sitter space, and the
basic relation \basicgflux\ leads to the right entropy from the
low energy open string description.

As we have discussed, this work raises many interesting questions
for future work.  One basic question is the counting of strings
stretched from the D-branes to the horizon (and in the global
geometry then stretching back to a causally disconnected region of
the D-brane wall).  This calculation, and its analogue in ordinary
AdS/CFT and in more generic AdS flux compactifications, would
provide a microscopic computation of the derivative of the entropy
with respect to flux quantum numbers. Another major question is
the problem of further testing (and more precisely formulating)
the conjecture of an open string description of ${\cal O}_s$'s
causal patch.  We plan to pursue this using the limit we discussed
in which the D-branes approach the horizon. Similarly, finding a
way to determine the precise relation between dS and DS entropy
(perhaps by pushing further along the lines discussed in \S5.6\
and \S7, or by studying black holes in D-Sitter space)
is crucial for further progress. Given a clear relation
between D-Sitter and de Sitter vacua, one may be able to study
other aspects of de Sitter physics using the D-Sitter degrees of
freedom.  For example, one may be able to study the distribution
and decay dynamics of flux backgrounds
\refs{\BoussoXA,\MaloneyRR,\KachruAW,\SusskindKW,\DouglasUM\freyetal}\
using our D-brane description, or elucidate some of
the conceptual puzzles raised in e.g. \WittenKN\GoheerVF.

Our approach based on the deformation of flux vacua into
spacetimes containing D-brane bubbles can be applied to many more
situations. For example, the Susskind-Witten entropy of AdS flux
compactifications can be studied as a function of flux quantum
numbers, which leads to similar questions to those discussed here
regarding its interpretation in terms of D-branes.  In general,
any time D-brane bubbles can be extracted by deformation from a
spacetime, one can study as we did here their entropy and the
relation of the deformation to the original spacetime.

\vskip 1cm \centerline{\bf Acknowledgements} We would like to
thank A. Adams, T. Banks, R. Bousso, A. Frey, S. Hellerman, S.
Kachru, A. Karch, A. Maloney, E. Martinec, J. McGreevy, S.
Shenker, A. Strominger, L. Susskind, and C.Vafa for useful
discussions. The work of E.S. is supported in part by the
Israel-U.S. Binational Science Foundation. The work of M.F. and
E.S. is supported by the DOE under contract DE-AC03-76SF00515, by
the NSF under contract 9870115 and by the A.P. Sloan Foundation, and
that of M.F. by a Stanford Graduate Fellowship.

\listrefs

\end